\documentclass[10pt,journal,twocolumn,twoside]{IEEEtran} 

\usepackage{graphicx}
\usepackage{epstopdf}
\usepackage{float}
\usepackage{algorithmic}
\usepackage{array}
\usepackage{amsmath}
\usepackage{amssymb}
\usepackage{mdwmath}
\usepackage{mdwtab}
\usepackage{eqparbox}
\usepackage{stfloats}
\usepackage{fixltx2e}
\usepackage{hyperref}
\usepackage{cleveref}
\usepackage{booktabs}
\usepackage{multirow}
\usepackage{makecell}
\usepackage{subfigure}

\hypersetup{hypertex=true,
            colorlinks=true,
            linkcolor=blue,
            anchorcolor=blue,
            citecolor=blue}
\usepackage{cases} 

\usepackage{upgreek}

\usepackage{xcolor}
\usepackage{makecell}
\usepackage{amsmath}
\usepackage[boxed,ruled,commentsnumbered]{algorithm2e}
\usepackage{url}
\usepackage{cite}
\ifCLASSOPTIONcompsoc
\usepackage[caption=false,font=normalsize,labelfont=sf,textfont=sf]{subfig}
\else

\allowdisplaybreaks[4]

\newtheorem{corollary}{\bf Corollary}
\newtheorem{theorem}{\bf Theorem}

\newtheorem{lemma}{\bf Lemma}

\makeatletter

\renewcommand*{\@opargbegintheorem}[3]{\trivlist
      \item[\hskip \labelsep{\bfseries #1\ #2}] \textbf{(#3):}\ }
\makeatother

\begin{document}
\title {Spectral Efficiency-Aware  Codebook Design  for Task-Oriented Semantic Communications}
\author{Anbang Zhang, Shuaishuai Guo,~\IEEEmembership{Senior Member, IEEE}, Chenyuan Feng,~\IEEEmembership{Member, IEEE}, \\Shuai Liu,~\IEEEmembership{Senior Member, IEEE}, Hongyang Du,~\IEEEmembership{Member, IEEE}, and Geyong Min,~\IEEEmembership{Member, IEEE} 
\thanks{
Anbang Zhang, Shuaishuai Guo and Shuai Liu are with School of Control Science and Engineering, Shandong University, Jinan 250061, China (e-mail: 202234946@mail.sdu.edu.cn, shuaishuai\_guo@sdu.edu.cn, liushuai@sdu.edu.cn); Hongyang Du is with University of Hong Kong, China (email: duhy@eee.hku.hk); Chenyuan Feng and Geyong Min are with the Department of Computer Science, University of Exeter, U.K. (email: c.feng@exeter.ac.uk, g.min@exeter.ac.uk).


}
}
\maketitle

\begin{abstract} 
Digital task-oriented semantic communication (ToSC) aims to transmit only task-relevant information, significantly reducing communication overhead. Existing ToSC methods typically rely on learned codebooks to encode semantic features and map them to constellation symbols. However, these codebooks are often sparsely activated, resulting in low spectral efficiency and underutilization of channel capacity. This highlights a key challenge: how to design a codebook that not only supports task-specific inference but also approaches the theoretical limits of channel capacity. To address this challenge, we construct a spectral efficiency-aware codebook design framework that explicitly incorporates the codebook activation probability into the optimization process. Beyond maximizing task performance, we introduce the Wasserstein (WS) distance as a regularization metric to minimize the gap between the learned activation distribution and the optimal channel input distribution. Furthermore, we reinterpret WS theory from a generative perspective to align with the semantic nature of ToSC. Combining the above two aspects, we propose a WS-based adaptive hybrid distribution scheme, termed WS-DC, which learns compact, task-driven and channel-aware latent representations. Experimental results demonstrate that WS-DC not only outperforms existing approaches in inference accuracy but also significantly improves codebook efficiency, offering a promising direction toward capacity-approaching semantic communication systems.

\end{abstract}


\section{Introduction} 
\IEEEPARstart{A}{rtificial} intelligence (AI) applications are evolving from data-driven Internet models \cite{9606720, 9134426} to systems that emphasize interaction between intelligent agents and their environments \cite{9979702}. These interactions demand efficient transmission of environmental data and user intent \cite{10388354}, especially amid growing data volumes. This shift exposes limitations in traditional communication paradigms and underscores the need for more efficient solutions \cite{9663101}. Semantic communication \cite{10666082}, particularly ToSC, has emerged as a promising approach for 6G networks by focusing on transmitting task-relevant meanings rather than full bit-streams \cite{10183789, 10024766}. ToSC ensures that transmitted information directly supports decision-making and actionable outcomes \cite{10758370}, significantly reducing communication overhead, enhancing robustness to channel variations \cite{10639525}, and improving the reliability of AI systems. 
Unlike conventional methods that emphasize channel capacity, ToSC leverages edge intelligence to minimize communication costs, offering a unified framework \cite{10454584, 9653664, 10930398} for data generation, transmission, and task execution.

\subsection{Related Works}
ToSC emphasizes accurately conveying the intended meaning of transmitted symbols \cite{10422886}. This enables system efficiency to be enhanced at the semantic layer, based on task requirements, rather than at the bit level. As a novel architecture combining task intelligence with data coding/decoding, ToSC has significant applications in task execution. For example, in data reconstruction tasks, \cite{9066966} introduces an autoencoder-based Joint Source-Channel Coding (JSCC) scheme that utilizes channel output feedback to improve receiver-side reconstruction quality. Attention-based semantic communication systems \cite{9450827} enhance speech signal recovery accuracy, while in \cite{9791409}, semantic encoding combined with Hybrid Automatic Repeat reQuest (HARQ) reduces semantic transmission errors. Numerous studies on ToSC aim to improve remote or edge inference tasks, such as classification \cite{9782523}, detection \cite{10556058}, and image processing \cite{9953099}, with design principles applied to various tasks \cite{10217150, 9837474, 9959884}. 


In practical applications, ToSC leverages more efficient architectures for feature extraction, specifically deep learning-based JSCC schemes (DeepJSCC). It jointly trains transceiver via an end-to-end loss function, surpassing traditional source-channel-separated coding \cite{Shao2021LearningTC}. However, DeepJSCC generates continuous input symbols rather than discretized constellation symbols, making it incompatible with digital modulation-based systems and burdensome for resource-constrained transmitters. Moreover, discrete symbols are better suited for complex tasks such as inference \cite{10159007} and predictive learning \cite{10101778}, highlighting the need for ToSC frameworks to learn richer discrete representations for improved digital transmission and task efficiency.

Recent research addresses this by combining ToSC with Vector Quantization (VQ), introducing codebooks for better compatibility with digital systems and optimal constellation mapping. The codeword interaction in ToSC acts as shared knowledge between the transceiver, transmitting only relevant semantic information while eliminating redundancy. Digital modulation in ToSC \cite{10159007} encodes features into discrete representations for better inference utility, and \cite{10110357} introduces an explainable ToSC design to enhance transmission efficiency. Generative VQ models, utilizing deterministic encoder-decoder architectures, have been applied to various tasks such as image generation \cite{9578911}, cross-modal translation \cite{9879208}, and image recognition \cite{yu2022vectorquantized}. A masked VQ-VAE-enabled codebook \cite{10101778} encodes feature representations for multiple downstream tasks, and \cite{10483054} proposes a framework balancing utility, informativeness, and security while reducing transmitted data. The above research shows that the task-driven codebook design is more critical for performing numerous edge tasks.

\begin{figure*}[]
\centering
\includegraphics[width=0.88\linewidth]{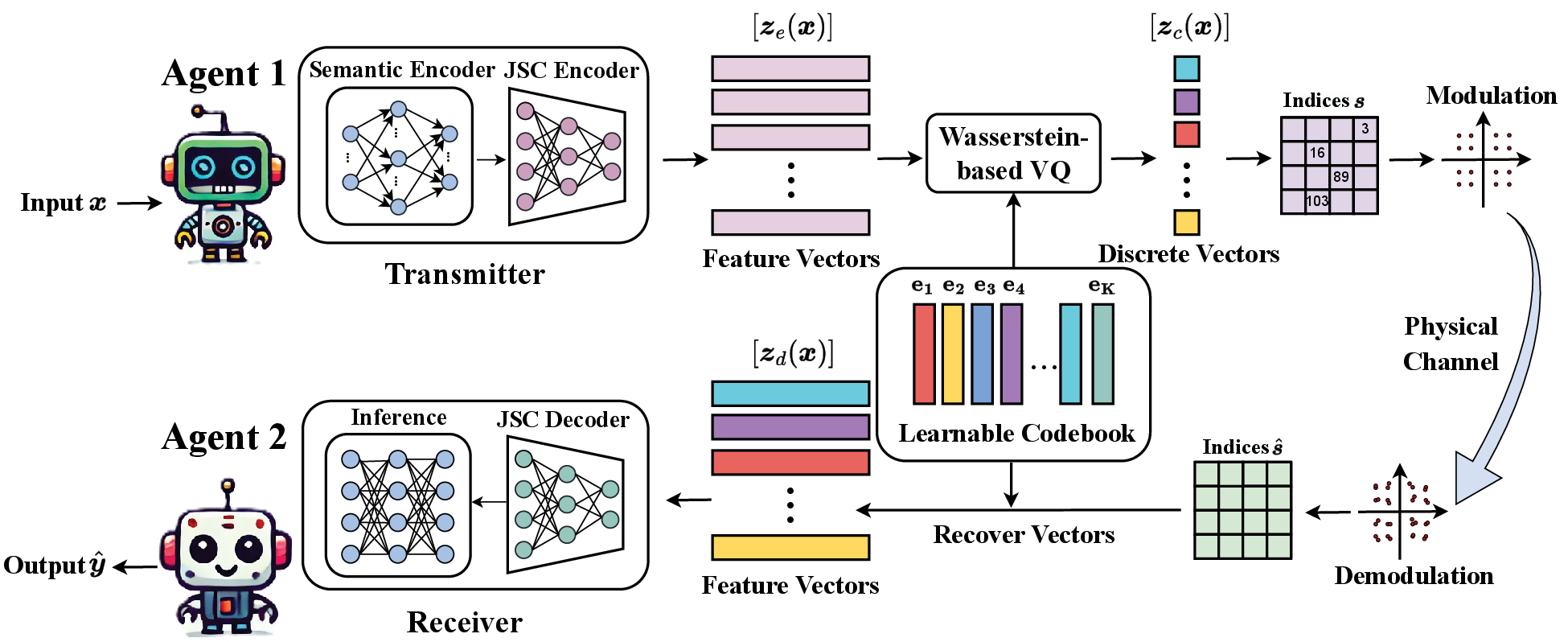}
\caption{Discrete task-oriented semantic communication considering codebook-modulation design.}
\label{fig1}
\end{figure*}

Note that the codebook design is sensitive to task characteristics, as codeword activation correlates with data distributions and task goals, which significantly impact ToSC performance. Following VQ-VAE, recent improvements aim to address codebook collapse (low utilization), where latent vectors are quantized into only a few codewords, losing semantic distinctions. Larger codebooks, though, often suffer from sparse clustering centers, making it difficult to capture up-to-date semantic content. To overcome this, various heuristics have been introduced, such as exponential moving average (EMA) updates \cite{razavi2019generating} and codebook resets \cite{3496104}. A scheme in \cite{pmlr-v162-takida22a} combines stochastic quantization and posterior classification distributions to extend VAE and improve codebook utilization. However, these heuristic approaches may fail to find the global optimal solution, limiting scalability and reliability in critical applications like medical diagnostics and autonomous vehicles. In high-dimensional spaces, codewords follow sparse distributions, and only a few are mapped to constellation symbols, limiting system efficiency. Many studies focus on optimizing the activation probability for better signal processing. For example, \cite{9252948} uses uniform quantization for DNN output mapping, while \cite{9450827, 9998051} address modulation issues with nonuniform quantization. Other methods \cite{10225385} replace the Sigmoid function with a ReLU activation. Despite improvements, sparse activation distributions still limit model effectiveness, and spectral efficiency is not fully optimized, leading to underutilized codewords. To address these challenges, ToSC must focus on adapting codebook activation to channel conditions, moving toward channel-aware designs.

\subsection{Motivations \& Contributions}
Driven by the challenges of task-driven and channel-aware codebook design, we propose an adaptive hybrid distribution strategy to promote balanced codeword utilization and smooth constellation symbol allocation, thereby enhancing spectral efficiency. 
The key contributions of this paper are as follows:
\begin{itemize}
\item To address sparse codebook activation and constellation symbols, we propose a spectral efficiency-aware design strategy. Using WS distance, our method aligns codebook activation with the optimal channel input distribution, enhancing spectral efficiency and channel awareness ability.
\item By highlighting the importance of aligning codebook design with the task needs, we extend Wasserstein theory to integrate task-specific goals from a
generative modeling perspective, thus ensuring that generated data aligns with among true distributions and task requirements. 
\item Considering the two aforementioned aspects, we propose a WS-based hybrid target matching scheme integrated into the learning process. By combining a uniform distribution with a smooth prior Gaussian, WS-DC encourages balanced codeword usage and channel adaptation. This design produces a compact, task-driven, and channel-aware codebook, thereby improving semantic expression. 
\item To validate the proposed WS-DC scheme, we conducted extensive image task experiments, using digital modulation with finite constellations. The results show that WS-DC outperforms baseline methods, improving inference utility with task-driven and channel-aware strategies. 
\end{itemize}

\section{System Model} 
\subsection{Digital ToSC System}
As shown in Fig. \ref{fig1}, we consider that the specific transceiver in a ToSC system is realized by co-designed and co-trained DNNs, with a shared trainable codebook space.
The detailed description is given below:

\subsubsection{Feature Encoding}
Assume the input is an image denoted by $\boldsymbol{x} \in \mathbb{R}^n$, with the corresponding target label $\boldsymbol{y}$. A feature extractor first derives task-relevant features from the raw input $\boldsymbol{x}$. These features are then passed through a joint source-channel (JSC) encoder, which maps them onto a continuous latent representation $\boldsymbol{z}_e(\boldsymbol{x}) \in \mathbb{R}^{Q \times D}$ within the latent space $\mathcal{Z}_e \subseteq \mathbb{R}^D$ as $ \boldsymbol{z}_{e}(\boldsymbol{x}) = T(\boldsymbol{x} ; \pmb{\theta})$, where ${T(\boldsymbol{x}; \pmb{\theta})}$ is a combination of the feature extractor and JSC encoder at the transmitter side with parameters $\pmb{\theta}$. 

\subsubsection{Discrete Codebook Mapping}
To ensure compatibility with digital communications, a VQ scheme is employed to discretize the continuous encoder outputs. Specifically, we define a learnable codebook $\mathcal{E} = {\mathbf{e}_k \in \mathbb{R}^D \mid k = 1, 2, \dots, K}$, where each basis vector $\mathbf{e}_k$ represents a codeword, and $K$ is the total number of codewords. The associated $Q$-dimensional discrete latent space is defined as the Cartesian power $\mathcal{E}^Q \subseteq \mathbb{R}^{Q \times D}$, where $Q$ denotes the number of components. Through VQ, the continuous latent vector $\boldsymbol{z}_e(\boldsymbol{x})$ is mapped to a discrete latent variable $\boldsymbol{z}_c(\boldsymbol{x}) \in \mathcal{E}^Q$, with its $q$-th component $\boldsymbol{z}_c^q(\boldsymbol{x}) \in \mathcal{E}$ for $q = 1, 2, \dots, Q$.
The mapping process is described by the following nearest-neighbor search method, as:
\begin{equation}\label{eq6}
\boldsymbol{z}_{c}^{q}(\boldsymbol{x})=\arg \min _{\mathbf{e}_{k}}\left\|\boldsymbol{z}_{e}^{q}(\boldsymbol{x})-\mathbf{e}_{k}\right\|_{2}, \forall \mathbf{e}_{k}\in \mathcal{E}.
\end{equation}

Regardless of the dimensionality of $\boldsymbol{z}_{c}(\boldsymbol{x})$, only the index of the selected codeword $\mathbf{e}_k$ is transmitted. Hence, the communication cost is limited to $\log_2 K$ bits per component.

\subsubsection{Digital Modulation and Demodulation}
Under a digital modulation scheme, the modulation process, physical channel, and demodulation together form an extended channel model that maps original codebook indices to their corrupted counterparts. Specifically, the codebook indices are sequentially mapped to constellation symbols. Let $\mathbf{s}$ denote the transmitted signal, which is corrupted by additive noise. The received symbols after demodulation can be expressed as $\hat{\boldsymbol{s}} = g(\boldsymbol{s} + \boldsymbol{n})$, where $g(\cdot)$ is the demodulation function and $\hat{\boldsymbol{s}}$ denotes the recovered symbols. For theoretical analysis, we adopt the widely used additive white Gaussian noise (AWGN) model, i.e., $\boldsymbol{n} \sim \mathcal{CN}(\mathbf{0}, \sigma^2 \mathbf{I})$. The receiver then performs a de-mapping operation by selecting the corresponding vectors $\boldsymbol{z}_d(\boldsymbol{x})$ from the discrete codebook based on $\hat{\boldsymbol{s}}$.

\subsubsection{Task Interference}
After demodulation, the discrete representation $\boldsymbol{z}{d}(\boldsymbol{x})$ is fed into the task inference module, formulated as $\hat{\boldsymbol{y}} = R(\boldsymbol{z}{d}(\boldsymbol{x}); \pmb{\eta})$, where $R(\cdot; \pmb{\eta})$ denotes a unified deep neural network (DNN) that includes the JSC decoder, with $\pmb{\eta}$ representing its trainable parameters.

Thus, this ToSC system is described by the  Markov chain:
\begin{equation}\label{eq4}
Y \rightarrow X \rightarrow Z_{e} \rightarrow Z_{c} \rightarrow S \rightarrow \hat{S} \rightarrow Z_{d} \rightarrow \hat{X} \rightarrow \hat{Y}, 
\end{equation}
satisfying 
\begin{align}\label{eq5}
P(\hat{\boldsymbol{y}} | \boldsymbol{x}) &= P_{\pmb{\eta}}(\hat{\boldsymbol{y}} | \boldsymbol{z}_{d}) 
P_d\left(\boldsymbol{z}_{d} | \hat{\boldsymbol{s}}\right) 
P_{\text{channel }}\left(\hat{\boldsymbol{s}} | \boldsymbol{s}\right) \\ \nonumber   
&\times P_m\left(\boldsymbol{s} | \boldsymbol{z}_{c} \right) 
P_e\left(\boldsymbol{z}_{c} | \boldsymbol{z}_{e}\right) 
P_{\pmb{\theta}}\left(\boldsymbol{z}_{e} | \boldsymbol{x}\right),
\end{align}
where $S$, $\hat{S}$, $Z_{e}$, $Z_{c}$, $Z_{d}$, $X$, $\hat{X}$ and $Y$ are random variables; $\boldsymbol{s}$, $\hat{\boldsymbol{s}}$, $\boldsymbol{z}_{e}$, $\boldsymbol{z}_{c}$, $\boldsymbol{z}_{d}$, $\boldsymbol{x}$, $\hat{\boldsymbol{x}}$, and $\mathbf{y}$ are their respective instances; and $P_{(\cdot)}(\cdot|\cdot)$  refers to the corresponding transmission function. 

\begin{figure}
        \centering
        \setlength{\belowcaptionskip}{-0.3Cm}   
        \subfigure[$\left|\boldsymbol{z}_{e}(\boldsymbol{x})\right| \gg\left|\mathcal{E}\right|$]{{\label{fig2a}}\includegraphics[width=0.49\linewidth]{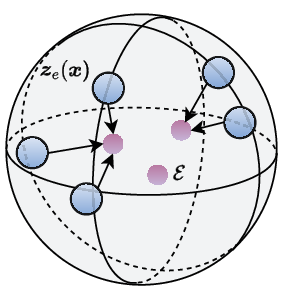}}          
        \subfigure[$\left|\boldsymbol{z}_{e}(\boldsymbol{x})\right| \ll\left|\mathcal{E}\right|$]{{\label{fig2b}}\includegraphics[width=0.49\linewidth]{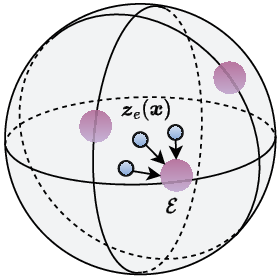}}
        \caption{The 3-D depiction of the effect of codeword utilization on discrete latent representations. (a) If the encoder outputs are larger, multiple codewords are likely to be used. (b) If the encoder outputs are smaller, they tend to cluster and map to fewer codewords.}
        \label{fig2}
\end{figure}

\begin{figure}
        \centering     \includegraphics[width=1\linewidth]{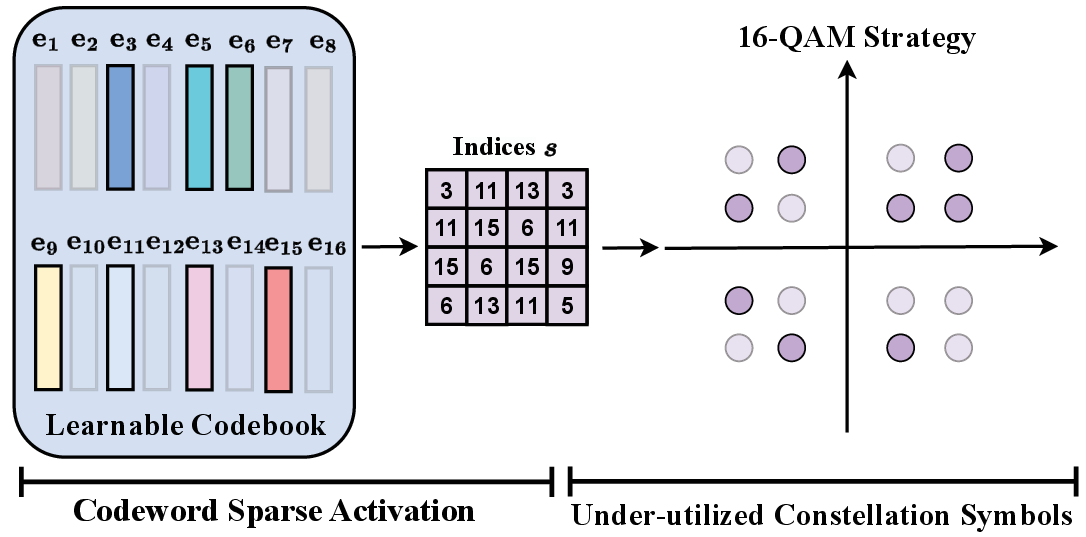}
        \caption{Index Collapse on Constellation Symbols with 16-QAM Strategy. In the case of 16-QAM, sparse activation in the codebook results in the activation of only a small subset of codewords. This causes a "collapse" in the constellation map, leading to suboptimal utilization of the signal space.}
        \label{fig3}
\end{figure}

\subsection{Problem Description}
In digital ToSC systems, the VQ scheme is typically used for discrete feature mapping. However, VQ processing is inherently prone to codebook collapse, which introduces task- and channel-aware challenges, as detailed below.

\subsubsection{Localized Pitfalls in VQ Space}
The dynamics of VQ training are illustrated in Fig. \ref{fig2}. The nearest-neighbor search method, commonly used in traditional VQ architectures, follows a greedy strategy. This approach typically results in the selection of only a few peripheral codewords, leaving the majority of codewords unused. Consequently, it is susceptible to getting stuck in suboptimal local solutions. Specifically, the sparse activation of the codebook leads to low codeword utilization. Such chronic under-utilization impedes the learning of rich data representations and task-specific goals. With limited information passing through the bottlenecks, the discrete latent representations fail to improve downstream task performance.

\subsubsection{Index Collapse about Constellation Symbols} 
The interaction between sparse codebook activation and constellation symbol mapping is depicted in Fig. \ref{fig3}. The sparse activation of the codebook leads to the concentration of mapped indices on a small subset of codewords, restricting the range of constellation symbols. As a result, many of the constellation symbols that should be utilized remain ineffective, causing a "collapse" in the constellation map, where the full signal space is not leveraged. This concentration also means that the available signal choices are severely limited, resulting in low spectral efficiency, poor channel utilization, and redundant signal transmission, ultimately reducing transmission efficiency.

\section{Enhanced ToSC with Channel-Aware Codebook Design} 
In this section, we investigate how to utilize WS distance to optimize the relationship between codebook activation and channel capacity in ToSC systems.

\subsection{Input Distribution Design for Efficient Channel Utilization}
According to the channel capacity theorem, the capacity $C$ is defined as $C=\max _{P_S(\boldsymbol{s})} I(S ; \hat{S})$, where $I(S ; \hat{S})$ is the mutual information between the input $S$ and output $\hat{S}$, and $P_S(\boldsymbol{s})$ is the probability density function, expressed as:
\begin{equation}\label{eq9}
\begin{aligned}
I(S ; \hat{S})=H(\hat{S})-H(\hat{S}|S),
\end{aligned}
\end{equation} 
where $H(\hat{S})$ is the differential entropy of $\hat{S}$ and $H(\hat{S}|S)$ is the the conditional differential entropy of $\hat{S}$ when $S$ is given. Maximizing MI can improves channel utilization by enhancing the input-output dependence.
Then, $H(\hat{S})$ is defined as:
\begin{equation}\label{eq10}
\begin{aligned}
H(\hat{S})=-\sum_{\hat{\boldsymbol{s}}\in\mathcal{\hat{S}}}P(\hat{\boldsymbol{s}})\log_2P(\hat{\boldsymbol{s}}),
\end{aligned}
\end{equation} 
where $\mathcal{\hat{S}}$ is the set on possible values of the random variable $\hat{S}$. Then, the conditional entropy $H(\hat{S}|S)$ is defined as:
\begin{equation}\label{eq11}
\begin{aligned}
H(\hat{S}|S)=-\sum_{\boldsymbol{s}\in\mathcal{S}}P(\boldsymbol{s})\sum_{\hat{\boldsymbol{s}}\in\mathcal{\hat{S}}}P(\hat{\boldsymbol{s}}|\boldsymbol{s})\log_2P(\hat{\boldsymbol{s}}|\boldsymbol{s}),
\end{aligned}
\end{equation} 
where $\mathcal{S}$ is the set on possible values of the random variable $S$. $H(\hat{S}|S)$ measures the remaining uncertainty in the random variable $\hat{S}$ given that the random variable $S$ is known.

Classical information theory shows that channel capacity hinges on the choice of input distribution, which must match the channel. For AWGN channels with an average‑power constraint, a Gaussian input is capacity‑achieving; in Rayleigh or Rician fading channels with channel state information (CSI), the optimum remains conditionally Gaussian. As a result, practical systems often adopt Gaussian or quasi‑Gaussian priors, which balance theoretical optimality with implement-ability and support distribution matching, codebook design, and performance optimization.

In practice, we elaborate with a specific AWGN. Given $S$, the conditional probability distribution of $\hat{S}$ remains constant, which is determined by a Gaussian noise distribution.
Thus, we can obtain:
\begin{equation}\label{eq12}
\begin{aligned}
P(\hat{\boldsymbol{s}}|\boldsymbol{s})=\frac{1}{\sqrt{2 \pi \sigma^{2}}} \exp \left(-\frac{(\hat{\boldsymbol{s}}-\boldsymbol{s})^{2}}{2 \sigma^{2}}\right).
\end{aligned}
\end{equation} 

Combining Eq.(\ref{eq11}) with Eq.(\ref{eq12}), we can obtain:
\begin{equation}\label{eq13}
\begin{aligned}
H(\hat{S}|S) = -\sum_{\boldsymbol{s} \in \mathcal{S}} & P(\boldsymbol{s}) \sum_{\hat{\boldsymbol{s}} \in \hat{\mathcal{S}}} \frac{1}{\sqrt{2 \pi \sigma^{2}}} \exp \left(-\frac{(\hat{\boldsymbol{s}}-\boldsymbol{s})^{2}}{2 \sigma^{2}}\right)\\
& \cdot \log _{2}\left(\frac{1}{\sqrt{2 \pi \sigma^{2}}} \exp \left(-\frac{(\hat{\boldsymbol{s}}-\boldsymbol{s})^{2}}{2 \sigma^{2}}\right)\right).
\end{aligned}
\end{equation}

Thus, based on \cite{cover2006elements}, we obtain the final conditional entropy as
$H(\hat{S}|S)=\frac{1}{2} \log _{2}\left(2 \pi e \sigma^{2}\right)$.
Then, the probability of the random variable $\hat{S}$ in Eq.(\ref{eq10}) can be expressed as follows:
\begin{equation}\label{eq18}
\begin{aligned}
P(\hat{\boldsymbol{s}}) = \sum_{\boldsymbol{s} \in \mathcal{S}} P(\boldsymbol{s}) P(\hat{\boldsymbol{s}}|\boldsymbol{s}).
\end{aligned}
\end{equation}

Combining Eq.(\ref{eq10}) and Eq.(\ref{eq18}), $H(\hat{S})$ is refined as:
\begin{equation}\label{eq19}
\begin{aligned}
H(\hat{S})=-\sum_{\hat{\boldsymbol{s}} \in \mathcal{S}}\left(\sum_{\boldsymbol{s} \in \mathcal{S}} P(\boldsymbol{s}) P(\hat{\boldsymbol{s}}|\boldsymbol{s})\right) \log _{2}\left(\sum_{\boldsymbol{s} \in \mathcal{S}} P(\boldsymbol{s}) P(\hat{\boldsymbol{s}}|\boldsymbol{s})\right).
\end{aligned}
\end{equation}

Note that, to maximize the channel capacity, we can choose an optimal prior distribution that maximizes $H(\hat{S})$. Based on the principle on maximum entropy, a Gaussian distribution has maximum entropy, given a power constraint. Assuming that the input signal $S$ satisfies a certain distribution $P_S(\boldsymbol{s})$, we know that the power constraint of input signal $S$ is $\mathbb{E}\left[S^{2}\right]=P_{*}$, where $P_{*}$ is the power of the input signal. Then, we can obtain:
\begin{equation}\label{eq21}
\begin{aligned}
P(\boldsymbol{s})=\frac{1}{\sqrt{2 \pi P_{*}}} \exp \left(-\frac{\hat{\boldsymbol{s}}^{2}}{2 P_{*}}\right).
\end{aligned}
\end{equation} 

Combining Eq.(\ref{eq12}) and Eq.(\ref{eq21}) to re-derive Eq.(\ref{eq18}), we get
\begin{equation}\label{eq22}
\begin{aligned}
P(\hat{\boldsymbol{s}})=
\frac{1}{\sqrt{2 \pi\left(P_{*}+\sigma^{2}\right)}} \exp \left(-\frac{\hat{\boldsymbol{s}}^{2}}{2\left(P_{*}+\sigma^{2}\right)}\right),
\end{aligned}
\end{equation}
where, $P(\hat{\boldsymbol{s}})$ remains a Gaussian distribution with mean $0$ and variance $P_{*}+\sigma^{2}$. Then, Eq.(\ref{eq19}) can be obtained as follow,
\begin{equation}\label{eq23}
\begin{aligned}
H(\hat{S})=
\frac{1}{2} \log _{2}\left(2 \pi e\left(P_{*}+\sigma^{2}\right)\right).
\end{aligned}
\end{equation}


After processing, we finally derive that the maximum channel capacity under Gaussian distribution as follows
\begin{equation}\label{eq25}
\begin{aligned}
C=&\max _{P_S(s)} I(S ; \hat{S})=\frac{1}{2} \log _{2}\left(1+\frac{P_{*}}{\sigma^{2}}\right).
\end{aligned}
\end{equation}
This derivation is fully consistent with the maximum channel capacity for Gaussian channels in information theory, showing that the channel capacity can reach the maximum transmission limit when the optimal channel input is Gaussian distribution.

\begin{figure*}
        \centering     \includegraphics[width=0.9\linewidth]{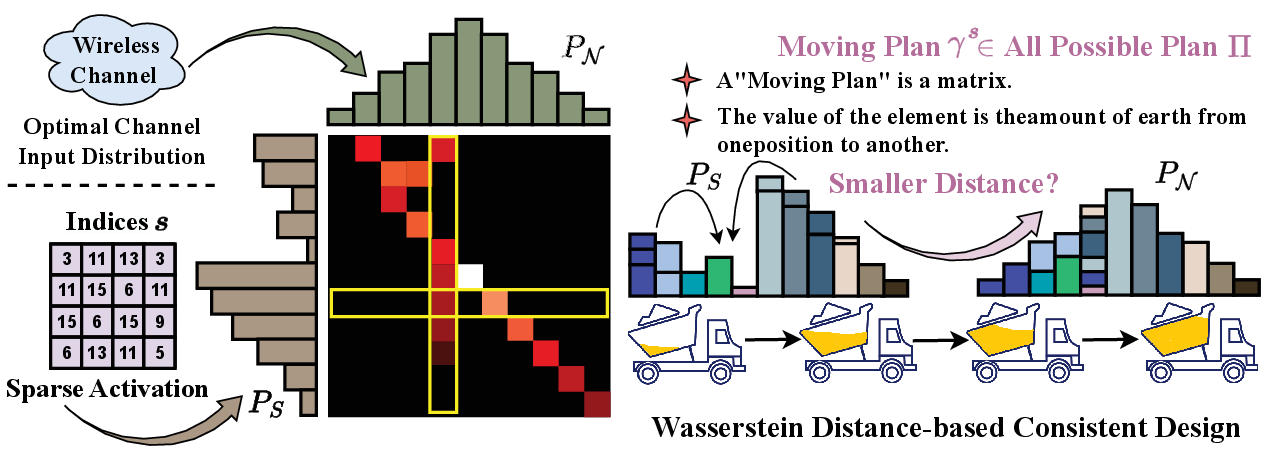}
        \caption{A distribution-consistent design scheme based on WS distance emphasizing the matching of codeword activation and optimal input distribution by optimizing the moving plan to enhance communication performance.}
        \label{fig4}
\end{figure*}

\subsection{WS-Based Codebook and Channel Input Alignment}



In modern digital ToSC systems, the transmitted signals are not directly transmitted using continuous input signals, but through discrete codebook indices $\boldsymbol{s}$. However, the activation probability distributions of these indices are determined by the task-relevant data and usually follow sparse distributions. Therefore, it is crucial to make the distributions of these activated codebook indices fit a optimal input distribution, thus improving the channel utilization efficiency. 

\textbf{\emph{WS distance-based strategy:}} The probability distribution with current known indices (i.e., channel inputs) is defined as $P_S(\boldsymbol{s})$, and the optimal input distribution in AWGN channel is Gaussian distribution $P_{\mathcal{N}}(\cdot)$. The goal is to make $P_S(\boldsymbol{s})$ closer to this target distribution $P_{\mathcal{N}}(\boldsymbol{s})$. Thus, assuming $\boldsymbol{s} \in\left\{\boldsymbol{s}_{1}, \boldsymbol{s}_{2}, \ldots, \boldsymbol{s}_{N_{\boldsymbol{s}}}\right\}$, WS distance is
\begin{equation}\label{eq26}
\begin{aligned}
\mathcal{W}\left(P_{S}, P_{\mathcal{N}}\right)=\inf _{\gamma^{\boldsymbol{s}} \in \Pi\left(P_{S}, P_{\mathcal{N}}\right)} \sum_{i=1}^{N_{\boldsymbol{s}}} \sum_{j=1}^{N_{\boldsymbol{s}}} \gamma^{\boldsymbol{s}}_{i j} \cdot d\left(\boldsymbol{s}_{i}, \boldsymbol{s}_{j}\right),
\end{aligned}
\end{equation}
where $\Pi\left(P_{S}, P_{\mathcal{N}}\right)$ denote the set of all joint distributions on $\boldsymbol{s}_{i}$ and $\boldsymbol{s}_{j}$, i.e., joint distributions satisfying the following marginal conditions $\gamma^{\boldsymbol{s}}$:
\begin{equation}\label{eq27}
\begin{aligned}
\sum_{j=1}^{N_{\boldsymbol{s}}} \gamma_{i j}=P_{S}\left(\boldsymbol{s}_{i}\right), \sum_{i=1}^{N} \gamma_{i j}=P_{N_{\boldsymbol{s}}}\left(\boldsymbol{s}_{j}\right), \gamma_{i j} \geq 0,
\end{aligned}
\end{equation}
where $\gamma^{\boldsymbol{s}}_{i j}$ is expressed as the mass transportation quantity from a point $\boldsymbol{s}_{i}$ in distribution $P_{S}$ to a point $\boldsymbol{s}_{j}$ in distribution $P_{\mathcal{N}}$. $d\left(\boldsymbol{s}_{i}, \boldsymbol{s}_{j}\right)$ is the distance measure, generally taken as Eq.(\ref{eq9}).


Ultimately, our goal is to optimize $P_{S}$ such that the WS distance is minimized, denoted as:
\begin{equation}\label{eq28}
\begin{aligned}
\min _{P_{S}(\boldsymbol{s})} \mathcal{W}\left(P_{S}, P_{\mathcal{N}}\right).
\end{aligned}
\end{equation}

As shown in Fig. \ref{fig4}, the WS distance directly transports one distribution to another, achieving the minimum total cost and not falling into a local optimum pitfalls. The aim is to match the codebook activation probability with the optimal input distribution, thus maximizing channel capacity utilization.


\section{Compact Discrete Semantic Coding: A Task-Driven Perspective} 
In this section, we develop a solid theoretical framework by connecting theory viewpoints about WS distance, VQ models, and deep discrete semantics in a task-driven perspective. 

\subsection{Wasserstein Optimization in Integrating Task Needs}

Firstly, we consider the training and application of the model throughout the end-to-end ToSC system. For the above training dataset $\boldsymbol{x}=\left\{x_{1}, \ldots, x_{N}\right\} \in {\mathbb{R}^{n}}$, we wish to learn a set of more compact codewords $\mathcal{E}=\left\{\mathbf{e}_{k}\in \mathbb{R}^{D}| k=1,2,\cdots,K\right\}$ on the coded latent space and an encoder to map each data examples into a latent space sequence $\mathcal{Z}_e\in {\mathbb{R}^D}$, preserving the deep features carried in the data. Then, we assign $Q$ a discrete distribution:
\begin{equation}\label{eq29}
\mathbb{P}_{\mathcal{E}, \pi^{q}}=\sum_{k=1}^{K} \pi_{k}^{q} \delta_{\mathbf{e}_{k}}, q=1, \ldots, Q,
\end{equation}
with the Dirac delta function $\delta$ and and the weights $\pi_{k}^{q} \in \Delta_{K-1}=\left\{\alpha \geq \mathbf{0}:\|\alpha\|_{1}=1\right\}$ in the ($K-1$)-simplex.

We first endow $Q$ discrete distributions over
$\mathcal{E}^{1}, \ldots, \mathcal{E}^{Q}$, sharing a common support set as the set of codewords $\mathcal{E}$. Then, we denote $\Gamma=\Gamma\left(\mathbb{P}_{\mathcal{E}, \pi^{1}}, \ldots, \mathbb{P}_{\mathcal{E}, \pi^{Q}}\right)$ to be the set of all joint distributions $\gamma \in \Gamma$, admitting these discrete distributions over $\mathcal{E}^{1}, \ldots, \mathcal{E}^{Q}$ as its marginal distribution ($\mathbb{P}_{\mathcal{E}, \pi^{1}}, \ldots, \mathbb{P}_{\mathcal{E}, \pi^{Q}}$) to sample a sequence of $Q$ codewords in $\mathcal{E}^{Q}$, $\left[\mathbf{\mathcal{E}}^{1}, \ldots, \mathbf{\mathcal{E}}^{Q}\right]$ corresponding to $Q$ latent codes $\boldsymbol{z}_{e}(\boldsymbol{x})$. Moreover, Let also define $\pi=\left[\pi^{1}, \ldots, \pi^{Q}\right]$ as the set of all weights. Note that a method in Eq. (\ref{eq5}) is prone to falling into local optimal pitfall, especially in high-dimensional and complex latent spaces, which may lead to suboptimal codeword assignments that do not align with the global objective.




To be end, based on generative viewpoint, we propose to learn the decoding part of the task-inference module, i.e., $f_{\text{TI}}: \boldsymbol{z}_{e}(\boldsymbol{x})\rightarrow \boldsymbol{x}/\boldsymbol{y}$ ( i.e., mapping from variable with the latent space $\mathcal{Z}_e$ to the original data space include label set), the codebook $\mathcal{E}$, and the weights $\pi$, to minimize:
\begin{equation}\label{eq30}
\min _{\mathcal{E}, \pi} \min _{\gamma \in \Gamma} \min _{f_{\text{TI}}} \mathcal{W}_{f_{d_{\boldsymbol{x}}}}\left(f_{\text{TI}} \# \gamma, \mathbb{P}_{\boldsymbol{x}}\right),
\end{equation}
where $\mathbb{P}_{\boldsymbol{x}}=\frac{1}{N} \sum_{n=1}^{N} \delta_{x_{n}}$ is the empirical data distribution and $f_{d_{\boldsymbol{x}}}$ is a metric function on the data space.

With respect to optimization problem (OP) in Eq.(\ref{eq30}), we give the same discrete distribution definition as above. The decoder $f_{\text{TI}}$ in task inference module is employed to map the codeword sequences to the data space and then complete the corresponding downstream tasks (e.g., classification tasks).

Subsequently, we learn $f_{\text{TI}}$ to minimize the codebook-data distortion given $\gamma$ to support the task inference, and finally adjust the codebooks $\mathcal{E}$, $\pi$ and $\gamma$ to minimize the optimal codebook-data distortion. To enrich the OP formulation, we introduce the Lemma \ref{Lemma1} to refresh the optimization process.

\begin{lemma}\label{Lemma1}
The optimal solution of the OP in Eq.(\ref{eq30}) is set as $\mathcal{E}^{*}=\left\{\mathbf{e}_{k}^{*}\right\}_{k}, \pi^{*}, \gamma^{*}$ and $f_{\text{TI}}$. Assuming that all combinations of codewords $K^{Q}<N$, then $\mathcal{E}^{*}=\left\{\mathbf{e}_{k}^{*}\right\}_{k}, \pi^{*}$ and $f_{\text{TI}}$ are also the optimal solution of the following OP:
\begin{equation}\label{eq31}
\min _{f_{\text{TI}}} \min _{\pi} \min _{\sigma_{1: Q} \in \Sigma_{\pi}} \sum_{i=1}^{N} f_{d_{\boldsymbol{x}}}\left(f_{\text{TI}}\left(\left[\mathbf{e}_{\sigma_{q}(i)}\right]_{q=1}^{Q}\right), x_{i}\right),
\end{equation}
where $\Sigma_{\pi}$ is the set of assignment functions
$\sigma$: $\{1, \ldots, N\} \rightarrow\{1, \ldots, K\}$. Here we denote $\sigma_{q}^{-1}(k)=\left\{i \in[N]: \sigma_{q}(i)=k\right\}$, $k=1,2,\cdots,K$ as the set of all data points assigned to the codeword $\mathbf{e}_{k}$, and $|\sigma_{q}^{-1}(k)|$ are proportional to $\pi_{k}^{q}$, which ensures that the usage frequency of each codeword is proportional to its weight, thus rendering the OP rational.
\end{lemma}

Then, we provide the proof in Appendix A. It can be appreciated that Lemma \ref{Lemma1} states that for the optimal solution in Eq.(\ref{eq30}), each data point $x_{i}$ is assigned to a prime $f_{TI}^{*}\left([\mathbf{e}_{\sigma_{q}^{*}}(i)]_{q=1}^{Q}\right)$ forming the optimal clustering centroids for the optimal clustering solution, minimizing map distortion.

Subject to the task needs, we incorporate a task loss term and perform a re-derivation. 
Then, we establish the following generalized theorem to engage the latent space with specific task loss term.

\begin{theorem}\label{Theorem1}
In order to transform the original OP into a more intuitive and tractable form, We can equivalently turn the optimization problem in Eq.(\ref{eq30}) as
\begin{equation}\label{eq32}
\min _{\mathcal{E}, \pi, \bar{f_{\text{TI}}}} \min _{\gamma \in \Gamma} \min _{\bar{T}: \bar{T} \# \mathbb{P}_{\boldsymbol{x}} = \gamma} \mathbb{E}_{(\boldsymbol{x},\boldsymbol{y}) \sim \mathbb{P}_{(\boldsymbol{x},\boldsymbol{y})}}\left[L_{\text {task }}\left(\bar{f_{\text{TI}}}\left(\bar{T}(\boldsymbol{x})\right); \boldsymbol{x},\boldsymbol{y}\right)\right],
\end{equation}
where $\bar{T}$ is modelled as a deterministic discrete encoder, which corresponds to the entire mapping process from data samples to $\boldsymbol{x}$ directly to a sequence of $Q$ codewords in $\mathcal{E}^{Q}$. And $\bar{f_{\text{TI}}}$ is a generalized task inference module, such as classification tasks, which also include a classifier.
\end{theorem}

The proof of Theorem \ref{Theorem1} is shown in Appendix B.
In details, Theorem \ref{Theorem1} learns firstly  codebook $\mathcal{E}$ and weights $\pi$. Next, the joint distribution $\gamma \in \Gamma$ is used to glue the codebook distributions. Subsequently, a deterministic discrete encoder is sought that maps the data samples to $\boldsymbol{x}$ directly to a sequence of $Q$ codewords extracted from $\gamma$, consistent with vector quantization and for further derivation. 
To make it trainable, we replace $\bar{T}$ by a continuous encoder $T(\boldsymbol{x} ; \pmb{\theta})$: $\boldsymbol{x}\rightarrow \boldsymbol{z}_{e}(\boldsymbol{x})$ in the following theorem.

\begin{theorem}\label{Theorem2}
If we seek $\bar{f}_{\text{TI}}$ and $\bar{T}$ in a family with infinite capacity (e.g., the family of all measurable functions), the object in Eq.(\ref{eq32}) are equivalent to following OP:
\begin{equation}\label{eq33}
\min _{\mathcal{E}, \pi,\gamma \in \Gamma} \min _{\bar{f}_{\text{TI}}, T}\left\{\begin{array}{c}
\mathbb{E}_{(\boldsymbol{x},\boldsymbol{y}) \sim \mathbb{P}_{(\boldsymbol{x},\boldsymbol{y})}}\left[L_{\text {task }}\left(f_{\text{TI}}\left({\text{VQ}}_{\mathcal{E}}\left(T(\boldsymbol{x})\right)\right); \boldsymbol{x}, \boldsymbol{y}\right)\right] \\
+\lambda \mathcal{W}_{f_{d_z}}\left(T \# \mathbb{P}_{\boldsymbol{x}}, \gamma\right)
\end{array}\right\},
\end{equation}
where ${\text{VQ}}_{\mathcal{E}}\left(f_{e}(x)\right)$ is a quantization operator, equivalent to Eq.(\ref{eq5}), which returns the closest codeword sequences to encoder output $T(\boldsymbol{x})$, and trade-off factor $\lambda > 0$. 
\end{theorem}

Here, we explain $f_{d_z}$ that given $\boldsymbol{z}_{e}(\boldsymbol{x}) = \left[\boldsymbol{z}_{e}(\boldsymbol{x})^{q}\right]_{q=1}^{Q}\in {\mathcal{Z}_e}$, $\boldsymbol{z}_{c}(\boldsymbol{x}) = \left[\boldsymbol{z}_{c}(\boldsymbol{x})^{q}\right]_{q=1}^{Q}\in \mathcal{E}^{Q}$, the distance is denoted as
\begin{equation}\label{eq34}
f_{d_{z}}(\boldsymbol{z}_{e}, \boldsymbol{z}_{c})=\frac{1}{q} \sum_{q=1}^{Q} \rho_{z}\left(\boldsymbol{z}_{e}^{q}, \boldsymbol{z}_{c}^{q}\right),
\end{equation}
where $\rho_{z}$ is a distance on the $q$-th component of $\boldsymbol{z}_{e}$ and $\boldsymbol{z}_{c}$.

The proof on Theorem \ref{Theorem2} is shown in Appendix C. It should be noted that in Theorem \ref{Theorem2}, we rigorously prove that the OPs of interest in Eq.(\ref{eq30}), Eq.(\ref{eq32}) and Eq.(\ref{eq33}) are equivalent under some mild conditions. This rationally explains why the OP in Eq.(\ref{eq33}) can be solved as our final tractable solution.

\subsection{Expressive Encoding by Compact Codeword Utilization}

Further, when considering Eq.(\ref{eq33}), we can develop a closed form of WS distance. Specifically, by minimizing the distance between latent representations before and after the discretisation, we can obtain a more compact task goal, i.e. minimizing $\mathcal{W}_{f_{d_z}}\left(T \# \mathbb{P}_{\boldsymbol{x}}, \gamma\right)$, where $\gamma$ admit $\mathbb{P}_{\mathcal{E}, \pi^{1: Q}}$ as its marginal distributions. Thus, we
implicitly minimize $\mathcal{W}_{\rho_{z}}\left(T^{q} \# \mathbb{P}_{\boldsymbol{x}}, \mathbb{P}_{\mathcal{E}, \pi^{q}}\right)$, since its upper bound is determined by the mean value of the marginal WS distance in each dimension.

\begin{lemma}\label{Lemma2}
To use codewords more efficiently, the WS distance $\min _{\pi} \min _{\gamma \in \Gamma} \mathcal{W}_{f_{d_{z}}}\left(T \# \mathbb{P}_{\boldsymbol{x}}, \gamma\right)$ is upper-bounded by
\begin{equation}\label{eq35}
\frac{1}{Q} \sum_{q=1}^{Q} \mathcal{W}_{\rho_{z}}\left(T^{q} \# \mathbb{P}_{\boldsymbol{x}}, \mathbb{P}_{\mathcal{E}, \pi^{q}}\right).
\end{equation}
\end{lemma}

According to Lemma \ref{Lemma2}, the OP in Eq.(\ref{eq33}) can be replaced by minimizing its upper-bound as follows
\begin{equation}\label{eq36}
\min _{\mathcal{E}, \pi} \min _{f_{\text{TI}}, T}\left\{\begin{array}{c}
\mathbb{E}_{(\boldsymbol{x},\boldsymbol{y}) \sim \mathbb{P}_{(\boldsymbol{x},\boldsymbol{y})}}\left[L_{\text {task }}\left(f_{\text{TI}}\left({\text{VQ}}_{\mathcal{E}}\left(T(\boldsymbol{x})\right)\right); \boldsymbol{x}, \boldsymbol{y}\right)\right] \\
+\frac{\lambda}{Q} \sum_{q=1}^{Q} \mathcal{W}_{\rho_{z}}\left(T^{q} \# \mathbb{P}_{\boldsymbol{x}}, \mathbb{P}_{\mathcal{E}, \pi^{q}}\right)
\end{array}\right\}.
\end{equation}

For the above formulations, we can adjust encoder $T$, task inference module $f_{\text{TI}}$, and codebook neng, so that the codewords can be more relevant to the task requirements and can be more expressive compared to the original data space. To explain the WS term in Eq.(\ref{eq36}), we make a reasonable demonstration using the Corollary \ref{corollary1}.

\begin{corollary}\label{corollary1}
Given $q \in[Q]$, consider minimizing the
term: $\min _{T, \mathcal{E}} \mathcal{W}_{\rho_{z}}\left(T^{q} \# \mathbb{P}_{\boldsymbol{x}}, \mathbb{P}_{\mathcal{E}, \pi^{q}}\right)$ in Eq.(\ref{eq36}). Thus, given $\pi^{q}$ and assume $K<N$, its optimal solution
$T^{*q}$ and $\mathcal{E}^{*}$ are also
the optimal solution of the OP:
\begin{equation}\label{eq37}
\min _{T, \mathcal{E}} \min _{\sigma \in \Sigma_{\pi}} \sum_{n=1}^{N} \rho_{z}\left(T^{q}\left(\boldsymbol{x}_{n}\right), \mathbf{e}_{\sigma(n)}\right)
\end{equation}
which is described in detail in Eq.(\ref{eq31}). Corollary 1 indicates the WS objective, which minimizes the codebook-latent distortion to make the codewords become the clustering centres of the latent representation $\boldsymbol{z}_{e}(\boldsymbol{x})$. Specially, our consideration is the optimal assignment function $\sigma^{*}$ at the optimal solution, which indicates how latent representations (or data examples) have more fitting properties to the clustering centroids (i.e., codewords), i.e., the cardinalities $|(\sigma_{q}^{*})^{-1}(k)|$, $k=1,2,\cdots,K$ are proportional to $\pi_{k}^{q}$.
\end{corollary}

\subsection{An Adaptive Hybrid Distribution Strategy}



Recall the key issues of VQ-based systems in section II, the sparse codeword activation leads to index collapse around constellation symbols, resulting in inefficient utilization of spectral resources. Inspired by the practical WS scheme in Eq.(\ref{eq36}), we introduce an adaptive hybrid distribution strategy that combines a Gaussian distribution with a uniform distribution to achieve both smoothness and diversity in codeword activation patterns. This yields two key benefits:
\begin{itemize}
\item We can encourage uniform codeword usage by shaping the learned codeword activation distribution $\pi^{q}$ toward a flat uniform distribution. This helps mitigate localized pitfalls and improves latent representation diversity.
\item We align the codeword activation distribution with the optimal channel input distribution (i.e. Gaussian distribution), enhancing the spectral and channel robustness of the transmitted representations. 
\end{itemize}


Specifically, we reset the weight in Eq.(\ref{eq36}) and describe the WS scheme as a hybrid target transport, defined as:
\begin{equation}
\mathcal{W}_{U} = \frac{1}{Q} \sum_{q=1}^{Q} \mathcal{W}_{\rho_{z}}\left(T^{q} \# \mathbb{P}_{\boldsymbol{x}}, \mathbb{P}_{\mathcal{E}, \pi^{q}}\right),
\end{equation}
where $\mathbb{P}_{\boldsymbol{x}}$ is designed to approximate a uniform distribution, and $\pi^{q}$ denotes the codeword activation distribution.

\begin{algorithm}[t]
  \label{alg1}
  \textbf{Initialize:} Encoder ${T(\boldsymbol{x}; \pmb{\theta})}$, Task Inference Module $R(\boldsymbol{z}_{d}(\boldsymbol{x}) ; \pmb{\eta })$, Codebook $\mathcal{E}$ and Codeword Weight Set $\left\{\pi^{q_c}=\operatorname{softmax}\left(\beta^{q_c}\right)\right\}_{q_c=1}^{Q_c}$.\\
  \KwIn{\textbf{Dataset} $\boldsymbol{x} \in \mathcal{X}$;}
  \For{\textbf{Training Rounds} epoch $= 0, 1, \dots$}{
  Randomly split dataset $\mathcal{X}$ into $B_m$ mini-batches;
    
  \For{\textbf{Each Batch} $iter = 1, \dots, B_m$}{
  Form an empirical batch distribution $\mathbb{P}_B$ based on the samples $\boldsymbol{x}_1,\dots,\boldsymbol{x}_B$;\\
  \emph{\textbf{Encoder:}} $E(\boldsymbol{x}_{1\dots B}; \pmb{\theta}) \rightarrow \boldsymbol{z}_{e}(\boldsymbol{x}_{1\dots B})$;\\

  \emph{\textbf{Discretization:}} ${\text{VQ}}_{\mathcal{E}}(\boldsymbol{z}_{e}(\boldsymbol{x}_{1\dots B})) \rightarrow \boldsymbol{z}_{c}(\boldsymbol{x}_{1\dots B})$;\\

  Process indices $k$ into constellation symbol $s$;\\
  
  \emph{\textbf{AWGN Channel:}} Transmit $s$ over the channel;

  \emph{\textbf{Task Inference:}} $R(\boldsymbol{z}_{d}(\boldsymbol{x}_{1\dots B}) ; \pmb{\eta})\rightarrow \hat{y}_{1\dots B}$;

 
   Update $T(\cdot;\pmb{\theta})$, $R(\cdot; \pmb{\eta})$, $\left\{\beta^{q_c}\right\}_{q_c=1}^{Q_C}$ and $\mathcal{E}$ by Eq.(\ref{eq38}).
    }
  }
  \KwOut{Trained $T( \cdot ; \pmb{\theta})$, $R( \cdot ; \pmb{\eta})$, and $\mathcal{E}$}
  \caption{WS Distance-Based Adaptive Strategy}
\end{algorithm}

Based on Eq.(\ref{eq26}), we further design the codeword activation distribution to align with a smoothed Gaussian target as:
\begin{equation}
\mathcal{W}_{G} = \frac{1}{Q} \sum_{q=1}^{Q} \mathcal{W}_{\rho_{z}}\left(\mathbb{P}_{\mathcal{N}}, \mathbb{P}_{\mathcal{E}, \pi^{q}}\right),
\end{equation}
where $\mathbb{P}_{\mathcal{N}},$ is the Gaussian target distribution. 

Combining $\mathcal{W}_{U}$ and $\mathcal{W}_{G}$,  we define a convex combination of the uniform distribution and the prior Gaussian, as follows:
\begin{equation}
\mathbb{P}_{H} = \alpha \mathbb{P}_{\boldsymbol{x}} + (1 - \alpha) P_{\mathcal{N}},
\end{equation}
where $P_{H}$ represents the final hybrid target distribution and $\alpha \in[0,1]$ is a weight coefficient that controls the trade-off between codeword diversity and smoothness. Then, we can obtain WS-based design scheme, defined as:
\begin{equation}
\begin{aligned}
\mathcal{W}_{H}=
\frac{1}{Q} \sum_{q=1}^{Q} \mathcal{W}_{\rho_{z}}\left(\alpha \mathbb{P}_{\boldsymbol{x}} + (1 - \alpha) P_{\mathcal{N}}, \mathbb{P}_{\mathcal{E}, \pi^{q}}\right) \\
\end{aligned}
\end{equation}

With the above theoretical derivation about codeword activation, we finally obtain a practical coding method as follow:
\begin{equation}\label{eq38}
\min\limits_{\mathcal{E},\pi,f_{\text{TI}},T}\begin{Bmatrix}\mathbb{E}_{(\boldsymbol{x},\boldsymbol{y}) \sim \mathbb{P}_{(\boldsymbol{x},\boldsymbol{y})}}\left[L_{\text {task }}\left(f_{\text{TI}}\left({\text{VQ}}_{\mathcal{E}}\left(T(\boldsymbol{x})\right)\right); \boldsymbol{x}, \boldsymbol{y}\right)\right]\\+\frac{1}{Q} \sum_{q=1}^{Q} \mathcal{W}_{\rho_{z}}\left(\alpha \mathbb{P}_{\boldsymbol{x}} + (1 - \alpha) P_{\mathcal{N}}, \mathbb{P}_{\mathcal{E}, \pi^{q}}\right),
\end{Bmatrix},
\end{equation}

For WS term, we parameterize $\pi^q = \pi^q(\beta^q)=\text{softmax}(\beta^q)$ with $\beta^q \in {\mathbb{R}^{K}}$. By aligning the codeword activation probabilities to this hybrid target, we ensure that the codewords are sufficiently utilized while adapting to channel characteristics. The pseudocode of WS-DC scheme is shown in Algorithm \ref{alg1}.

\section{Simulations \& Discussions} 


In this section, we compare the performance of WS-DC scheme with other DL-based baseline scheme. It is noteworthy that WS-DC is not specified for the image classification task.

\subsection{Experimental Settings}
\subsubsection{Compared Methods}
We evaluate WS-DC against three state-of-the-art ToSC baselines, described as follows:
\begin{itemize}
\item \textbf{DeepJSCC-G:} Based on NECST \cite{Choi2018NeuralJS}, this method utilizes a DNN to map input data to channel symbols and jointly learns both encoding and decoding processes. To distinguish the original NECST, we refer to it as DeepJSCC-G, and employ the cross-entropy loss for the training stages.
\item \textbf{DeepJSCC-VIB:} This method builds on VFE \cite{Shao2021LearningTC}, a learning-based ToSC framework that leverages the variational information bottleneck (VIB) principle to minimize communication cost.
Our design follows the same training strategy and is denoted as DeepJSCC-VIB for clarity.   
\item \textbf{DeepJSCC-RIB:} Derived from DT-JSCC \cite{10159007} scheme, this baseline method incorporates a robust information bottleneck (RIB) to enhance resilience against channel variations, achieving a balance between informativeness and robustness. We adopt the same training scheme and refer to this version as DeepJSCC-RIB.  
\end{itemize}

To ensure the fair comparisons, we standardize the experimental environment across all methods, including WS-DC. Specifically, we adopt a consistent discrete codebook design and apply the same modulation scheme $K$-ary quadrature amplitude modulation ($K$-QAM) to map discrete latent representations into channel symbols. Furthermore, all models share an identical neural network backbone, subject to equal constraints on computational complexity and memory usage, simulating deployment on resource-limited edge devices.

\begin{table}[t]
\setlength{\belowcaptionskip}{-60pt}   
\caption{Architecture Settings of Each Module}
\label{cifarstructure}
\centering
\renewcommand{\arraystretch}{1.03}
\begin{tabular}{@{}c|l|l@{}}
\toprule
\multicolumn{1}{l|}{}                                                    & \multicolumn{1}{c|}{\textbf{Layer}}                                                                                   & \multicolumn{1}{c}{\textbf{Outputs}}                                 \\ \midrule
\textbf{Encoder}                                                     & \begin{tabular}[c]{@{}l@{}}Conv$\times2$ + ResNet Block$\times1$ \\ Conv$\times4$ + ResNet Block$\times1$\end{tabular}              & \begin{tabular}[c]{@{}l@{}}$128\times16\times16$\\ $128\times4\times4$\end{tabular}      \\ \midrule
\textbf{\begin{tabular}[c]{@{}c@{}}Receiver\\ (Task inference module)\end{tabular}} & \begin{tabular}[c]{@{}l@{}}Conv$\times2$ + ResNet Block$\times2$\\ Maxpooling\\ Dense + Softmax\end{tabular} & \begin{tabular}[c]{@{}l@{}} $64\times4\times4$\\ $64$\\ $10$\end{tabular}  \\\bottomrule
\end{tabular}
\end{table}

\begin{figure*}[!t]
        \centering
        \subfigure[Codebook size $K=16$.]{{\label{fig5a}}\includegraphics[width=0.325\linewidth]{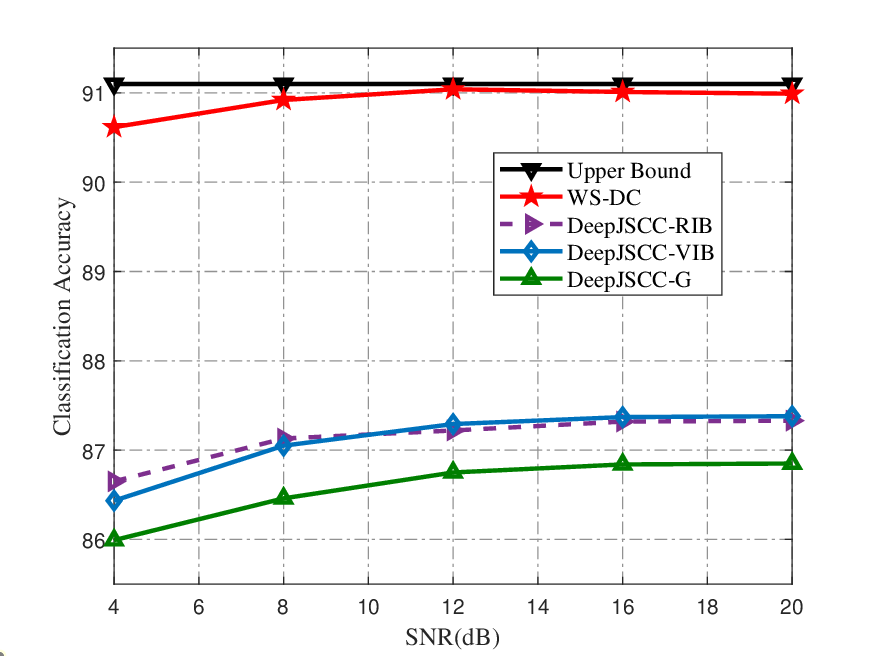}}
        \subfigure[Codebook size $K=64$.]{{\label{fig5b}}\includegraphics[width=0.32\linewidth]{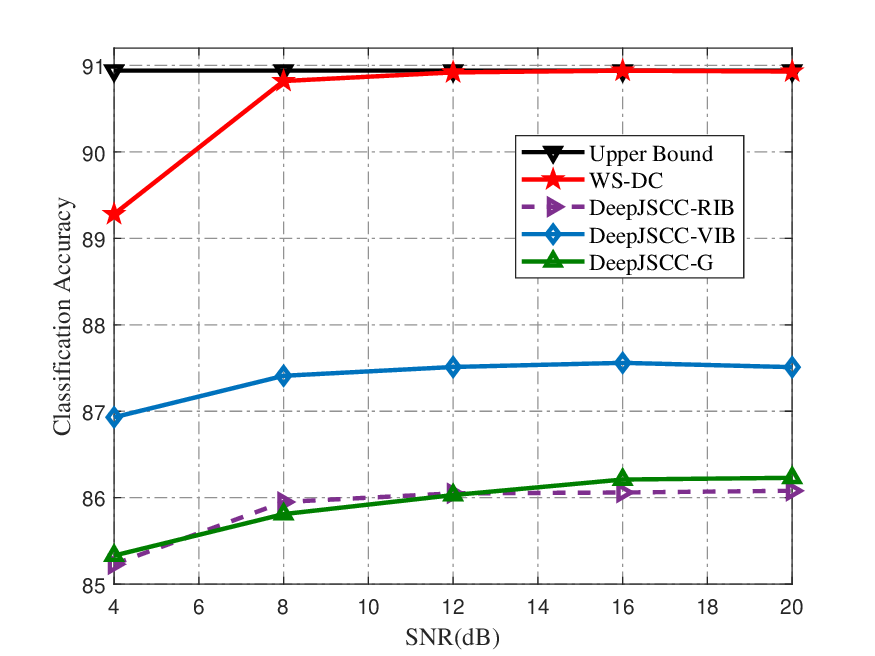}}
        \subfigure[Codebook size $K=256$.]{{\label{fig5c}}\includegraphics[width=0.32\linewidth]{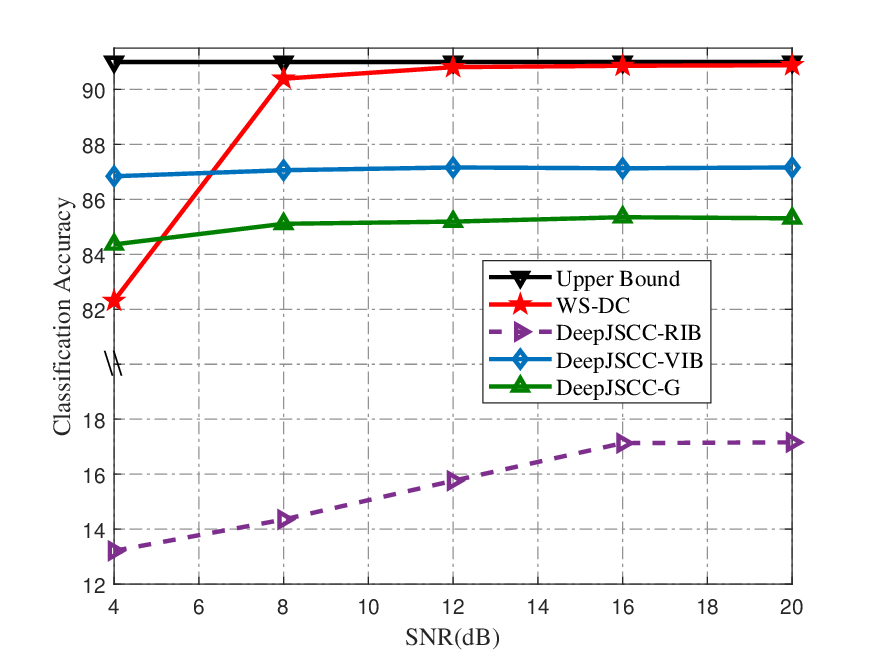}}
        \caption{Task performance comparison versus SNR, where Codebook size $K=16$, $K=64$, $K=256$ and Codeword dimension $D=64$.}
        \label{fig5}
\end{figure*}

\begin{figure*}[!t]
        \centering
        \subfigure[Codebook size $K=16$.]{{\label{fig6a}}\includegraphics[width=0.32\linewidth]{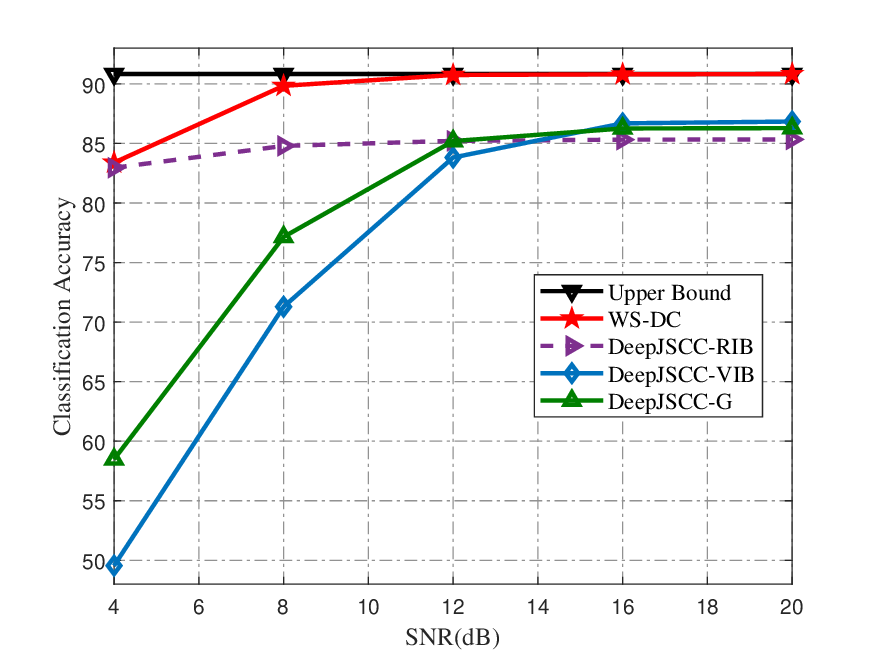}}
        \subfigure[Codebook size $K=64$.]{{\label{fig6b}}\includegraphics[width=0.32\linewidth]{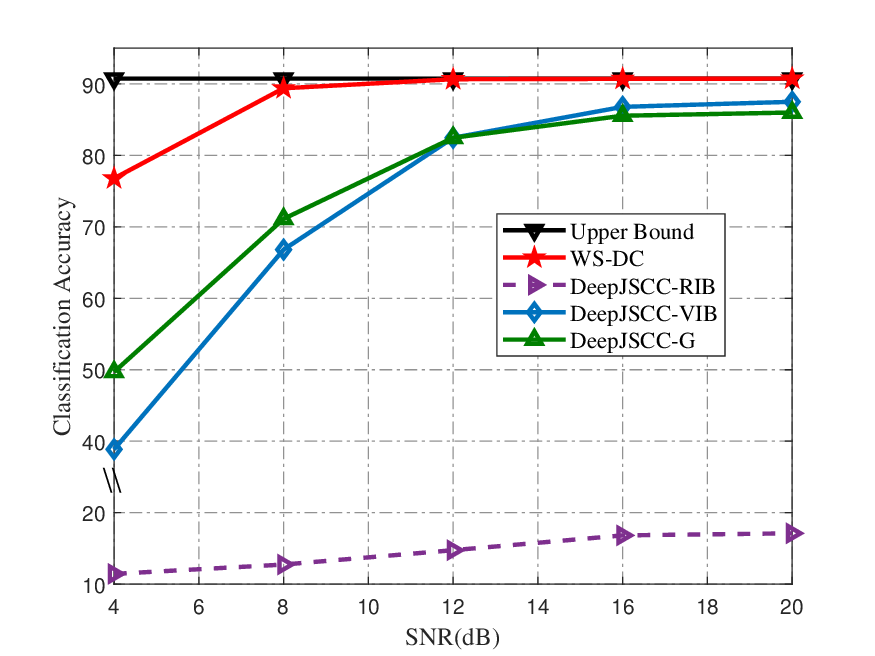}}
        \subfigure[Codebook size $K=256$.]{{\label{fig6c}}\includegraphics[width=0.32\linewidth]{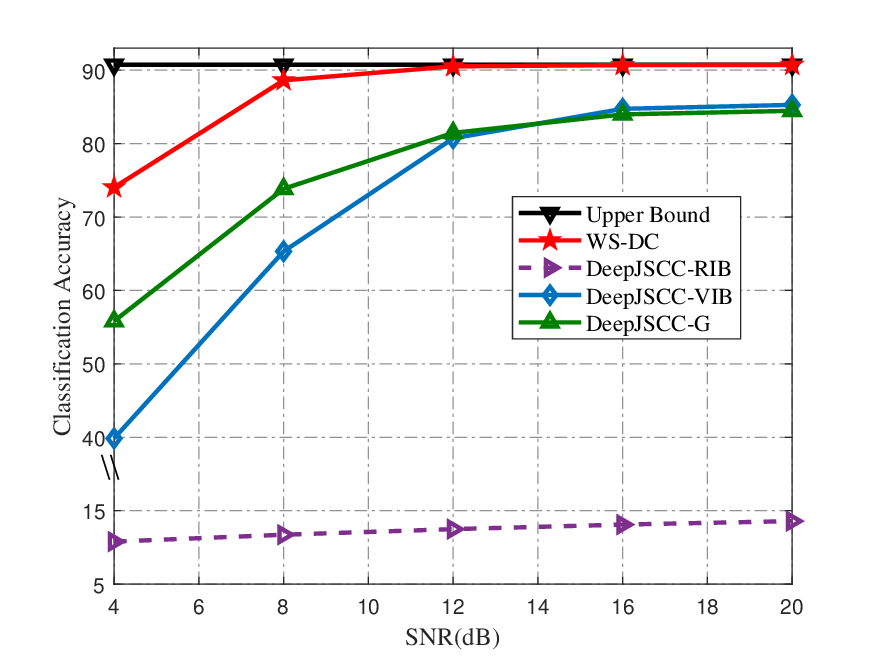}}
        \caption{Task performance comparison versus SNR, where Codebook size $K=16$, $K=64$, $K=256$ and Codeword dimension $D=512$.}
        \label{fig6}
\end{figure*}




\subsubsection{Dataset and Neural Network Architecture}
To evaluate performance, we utilize the CIFAR-10 dataset, which includes 50,000 training images and 10,000 test images. During training, we apply standard data augmentation techniques, including random cropping and horizontal flipping, to enhance generalization. Based on CIFAR-10, we design the model backbone using convolutional layers and ResNet blocks. The detailed architecture and parameter settings of the overall module in all compared methods are summarized in Tab. \ref{cifarstructure}.

\subsection{Experimental Results on Task Inference Performance}


In the experiments, we set three widely used codebook space sizes as $K=16$, $K=64$, and $K=256$, which correspond to $16$-QAM, $64$-QAM, and $256$-QAM modulation schemes. As for the codeword dimensions, $D=64$ and $D=512$ are set to explore how much they affects the actual performance. All the schemes are trained under $\operatorname{SNR}_{\text {train}}=12$dB and tested under $\operatorname{SNR}_{\text {test}} \in\{4 \mathrm{~dB}, 8 \mathrm{~dB}, 12 \mathrm{~dB}, 16 \mathrm{~dB}, 20 \mathrm{~dB}\}$, respectively. In this basic experiment, we fixed the target trade-off parameter $\alpha=0.5$ to maintain a balanced distribution.


\emph{1) Robustness Advantages of Compact Coding:} As shown in Fig. \ref{fig5a} and Fig. \ref{fig6a}, WS-DC exhibits excellent channel adaptation in the $K=16$ scenario. When $D=64$, its performance waves $90.62\%\sim90.99\%$ in the $4\mathrm{~dB}\sim20\mathrm{~dB}$, which is significantly better than DeepJSCC-RIB ($86.65\%\sim87.33\%$), DeepJSCC-VIB ($86.43\%\sim87.38\%$) and DeepJSCC-G ($85.99\%\sim86.85\%$). Even when $D=512$, WS-DC maintains $83.39\%$ under worse channel conditions ($4 \mathrm{~dB}$); and when the SNR $\ge 12 \mathrm{~dB}$, it still manages to reach $90.75\%\sim 90.83\%$, reaching the highest accuracy among all schemes. 
In contrast, other baseline schemes show sharp performance degradation at $4 \mathrm{~dB}$ (e.g., as low as $49.55\%$ for DeepJSCC-VIB and $58.48\%$ for DeepJSCC-G).  
As the codeword dimension increases, WS-DC remains stable, while other schemes decreases significantly. Even under low SNR conditions, WS-DC maximizes the key semantic information preservation and exhibits strong channel robustness.

\emph{2) Stability Challenges in Codebook Expansion:} As the codebook size expands, WS-DC remains above the best level, but ohter schemes shows task performance degradation in Fig. \ref{fig5b} and Fig. \ref{fig6b}. Specifically, when the codebook size is $K=64$, WS-DC under the regular SNR $\ge 12 \mathrm{~dB}$ is $90.92\%\sim 90.93\%$ at $D=64$ and $90.66\%\sim 90.74\%$ at $D=512$. And even at $4 \mathrm{~dB}$, it still maintains $89.28\%$ ($D=64$) and $76.78\%$ ($D=512$). In contrast, DeepJSCC-VIB and DeepJSCC-G, although their task performance eventually reaches about $87.5\%$ and $85.99\%$ ($D=512$) as the SNR increases, they fails to be satisfactory at $4 \mathrm{~dB}$, i.e., $38.86\%$ and $49.69\%$. Notably, DeepJSCC-RIB exhibits a disastrous result at $D=512$, and its utility is maintained at $11.43\%\sim17.11\%$. This is due to the difficulty of adapting the its training scheme to larger codebook spaces and dymantic channel conditions. 
The codebook expansion ($K=64$) enhances expressive capability, but redundant representations ($D=512$) exacerbate channel interference. Specially, at low SNR, DeepJSCC-RIB and DeepJSCC-G lead to a rapid accuracy degradation due to the lack of sufficient channel awareness. In contrast, WS-DC can minimize redundant information interference, thus maintaining high classification accuracy in this scenario.

\begin{table*}[t]
\centering
\caption{Performance comparison of WS-DC with varying codeword activation mixing factor $\alpha$, compared with other DeepJSCC baselines. Accuracy values under different SNR levels are also reported.}
\begin{tabular}{ccccccccc}
\toprule
\label{TAB2}
Model & Weight Init $\pi^{q}$ & $\alpha$ & OT Cost $\downarrow$ & Acc@4dB (\%) & Acc@8dB (\%) & Acc@12dB (\%) & Acc@16dB (\%) & Acc@20dB (\%) \\
\midrule
\multirow{6}{*}{WS-DC}
& \multirow{5}{*}{Uniform}
    & $0$ & 30.25 & 75.04 & 88.81 & 90.58 & 90.72 & 90.72\\
    & & $0.2$ & 24.19 & 75.24 & 88.98 & 90.74 & 90.83 & 90.86\\
    & & $0.4$ & 18.14 & 75.52 & 89.01 & 90.57 & 90.67 & 90.68\\
    & & $0.6$ & 12.09 & 73.39 & 88.07 & 90.17 & 90.35 & 90.36\\
    & & $0.8$ & 6.05 & 75.27 & 88.86 & 90.57 & 90.58 & 90.58\\
    & & $1$ & 0 & 57.38 & 83.19 & 89.94 & 90.28 & 90.31\\
\midrule
    DeepJSCC-G & Uniform & – & –  & 55.82  & 73.84 & 81.46 & 83.98 & 84.47\\
    DeepJSCC-VIB & Uniform & – & –  & 39.85  & 65.32 & 80.72 & 84.74 & 85.28\\
    DeepJSCC-RIB & Uniform & – & –  & 10.79  & 11.73 & 12.49 & 13.11 & 13.58\\
\bottomrule
\end{tabular}
\end{table*}

\emph{3) Systemic Advantages in High-dimensional Large-scale Codebook:} At extreme conditions $K=256$, the performance difference among all schemes is particularly significant, shown in Fig. \ref{fig5c} and Fig. \ref{fig6c}. when $D=64$, WS-DC still provides a higher task accuracy ($90.81\%\sim 90.88\%$) at SNR $\geq 12\mathrm{~dB}$. And when $D=512$, WS-DC fluctuates at $90.53\%\sim 90.71\%$. In contrast, DeepJSCC-RIB is extremely unstable in large codebook sizes. Regardless of codebook dimension, its accuracy fluctuates between $10\% \sim 20\%$, leading to a disastrous performance. In addition, DeepJSCC-VIB and DeepJSCC-G show a more moderate decline under low-dimensional conditions. They final performance maintains at a better level at SNR $\geq 12\mathrm{~dB}$ (i.e., $86.84\%\sim 87.16\%$ and $84.36\%\sim 85.35\%$). And at $4 \mathrm{~dB}$, DeepJSCC-VIB and DeepJSCC-G in high-dimensional transmission are slightly improved, compared to that at $K=64$.
The large codebook size brings stronger information expression capability to the model, but also makes the complexity of information compression and transmission increase significantly. 
In particular, at low SNR, the increased freedom may not bring better performance, but exacerbate the channel interference and overfitting problems. However, WS-DC significantly mitigates the overfitting and feature dispersion problems through channel-aware and task-driven strategy.

\begin{figure}[t]
\centering
\includegraphics[width=0.96\linewidth]{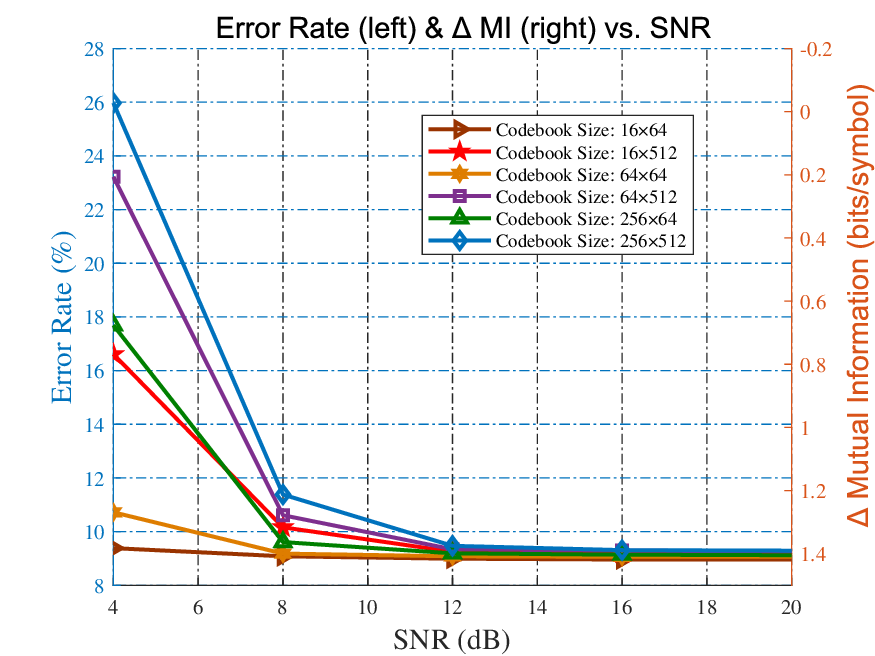}
\caption{Error Rate and $\Delta$MI vs. SNR for different codebook sizes. $\Delta$MI is defined as $\Delta \mathrm{MI} = I(Z_d; Y) - I(Z_e; Y)$, which measures the loss of task-relevant semantics information across the channel.}
\label{fig7}
\end{figure}

\subsection{Semantic Transmission Robustness Analysis}
We conducted experiments to investigate the impact of different codebook sizes on the preservation of task-relevant information under different SNR conditions. Specifically, we measure the degradation of task-relevant semantics in the channel through mutual information (MI) difference $\Delta$MI, where $\Delta$MI is defined as the difference between the mutual information between the encoder feature $Z_e$ and the label $Y$ and the mutual information between the received feature $Z_d$ and $Y$. As shown in Fig. \ref{fig7}, the results indicate that regardless of the SNR, our proposed WS-DC scheme achieves an enhancement of task-related information through the channel. For example, under low SNR conditions (e.g., $4 \mathrm{~dB}$ and $8 \mathrm{~dB}$), smaller codebooks (e.g., 16$\times$64) result in significantly higher $\Delta$MI values, indicating a greater increase in task-related information. With the significant increase in error rates, especially in the 256$\times$512 setting, the sensitivity of semantic representations to noise increases due to the larger quantization space and limited channel capacity. As the SNR improves, $\Delta$MI shows an upward trend and eventually converges across different settings, indicating enhanced channel matching for semantic transmission. Therefore, under channel-aware design, semantic transmission can better match spectrum-efficiency utilization, significantly improve the preservation of task-related information, and enhance channel robustness.

\subsection{Ablation Study}
In the ablation experiment, we compared the WS-DC performance under different mixed distribution weight factors $\alpha$ and OT costs, as well as with all DeepJSCC baselines. Specifically, we consider the most complex scenario, where the codebook size is $K=256$ and $D=512$, i.e. $256$-QAM, which has a more complex semantic representation.

Tab. \ref{TAB2} shows that as $\alpha$ increased from 0 to 1, the OT cost gradually decreased, indicating a more balanced utilization of codewords and a smoother activation distribution. Specifically, when $\alpha=0.6$, WS-DC achieved a relatively low OT cost of 12.09 while maintaining high accuracy at different SNR levels (e.g. 73.39\% at 4$\mathrm{~dB}$ and 90.36\% at 20$\mathrm{~dB}$). This indicates that an appropriate mixture of uniform prior and Gaussian prior effectively improves codebook efficiency and robustness to noise. However, displaying a single distribution in the first and sixth rows can have a certain impact on model performance. Reasonable weighting factors can make the codeword activation distribution more suitable for the original data and channel conditions. In addition, WS-DC consistently outperforms other baseline schemes, especially under low SNR conditions. For example, at 4$\mathrm{~dB}$, WS-DC achieved an accuracy of 75.24\% with $\alpha=0.2$, significantly higher than DeepJSCC-G (55.82\%) and DeepJSCC-VIB (39.85\%).
These results indicate that the proposed adaptive hybrid distribution strategy in WS-DC can achieve an excellent trade-off between transmission efficiency and task performance.

\begin{table}[t]
  \centering
  \caption{Computing complexity, Model complexity, Model Traing cost and Learning latency compared with other DeepJSCC baselines over Cifar-10 dataset.}
  \begin{tabular}{lccccc}
    \toprule
    \label{TAB3}
    Model & FLOPs & Params & Train (1x) & Test (1x) \\
    \midrule
    WS-DC & 0.17 G & 7.42 M & 35.91 s & 0.001 s\\
    DeepJSCC-G & 0.17 G & 7.42 M & 34.99 s & 0.001 s \\
    DeepJSCC-VIB & 0.17 G & 7.42 M & 28.81 s & 0.002 s \\
    DeepJSCC-RIB & 0.17 G & 7.42 M & 38.11 s & 0.001 s \\
    \bottomrule
  \end{tabular}
\end{table}

\subsection{Cost and Delay Comparison}
Tab. \ref{TAB3} shows the computational cost, complexity, and communication latency of all schemes running on an 11th Gen Intel(R) processor at 2.50 GHz and a single 3060 GPU core, including the specific number of floating-point operations (FLOPs) in the whole training process, the number of parameters in all the neural network model, the time for a single training epoch (batch-size: 512) and the task-inference time for a single image. 

The results indicate that all models have the same computational complexity and parameter scale, with each model having a FLOP of 0.17G and approximately 7.42M parameters, indicating the same computational requirements and maintaining lower complexity, as evidenced by their FLOPs and parameter counts. This is because the model structures under fair comparison are the same. However, the training cost of each round of WS-DC (35.91s) is slightly higher than that of DeepJSCC-G (34.99s) and DeepJSCC VIB (28.81s), mainly due to the introduction of Wasserstein based hybrid regularization. Nevertheless, WS-DC maintained a considerable testing latency (0.001s) compared to other baseline models, highlighting its efficiency in the inference process.
Thus, we can obtain that WS-DC can improve task performance and spectral efficiency by moderately increasing training costs without affecting inference speed. This further validates its practicality in real-time semantic communication scenarios.

\section{Conclusion}
This paper addresses the challenge of improving spectral efficiency and task performance in digital ToSC systems. We propose a WS-based adaptive hybrid distribution strategy that integrates both channel-aware and task-driven perspective, balancing uniform codeword utilization with alignment to the optimal channel input distribution. Moreover, we reformulate the generalized workflow theory from a generative perspective to better align with task-specific objectives. 
Experimental results demonstrate that the proposed WS-DC consistently outperforms existing methods across different codebook settings, establishing a more efficient and robust adaptive communication framework. 
In large-scale complex scenarios, WS-DC delivers high-quality task inference under bandwidth or channel constraints, thus advancing edge intelligence from data-driven to task-driven paradigms.

\begin{appendices}
\section{Proof of Lemma \ref{Lemma1}}
Regarding the more intuitive properties on Lemma \ref{Lemma1}, we mainly offer this generative process. Through the description in Section IV-A, we can address the different codebook conditions as follows. Firstly, $\gamma \in \Gamma$ is a distribution over $\mathcal{E}^{Q}$ with $\gamma\left(\left[\mathbf{e}_{i_{1}}, \ldots, \mathbf{e}_{i_{Q}}\right]\right)$, which satisfying $\sum_{i_{1}, \ldots, i_{q}=k, \ldots, i_{Q}} \gamma\left(\left[\mathbf{e}_{i_{1}}, \ldots, \mathbf{e}_{i_{Q}}\right]\right)=\pi_{k}^{q}$. And $f_{\text{TI}} \# \gamma$ is a distribution over $f_{\text{TI}}\left(\left[\mathbf{e}_{i_{1}}, \ldots, \mathbf{e}_{i_{Q}}\right]\right)$, with the mass $\gamma\left(\left[\mathbf{e}_{i_{1}}, \ldots, \mathbf{e}_{i_{Q}}\right]\right)$ or in other words, this implies that
\begin{equation}\label{eq44}
f_{\text{TI}} \# \gamma=\sum_{i_{1}, \ldots, i_{Q}} \gamma\left(\left[\mathbf{e}_{i_{1}}, \ldots, \mathbf{e}_{i_{Q}}\right]\right) \delta_{f_{\text{TI}}\left(\left[\mathbf{e}_{i_{1}}, \ldots, \mathbf{e}_{i_{Q}}\right]\right)}.
\end{equation}

Thus, we reach a preliminary OP from Eq.(\ref{eq30}):
\begin{equation}\label{eq45}
\begin{aligned}
\min _{\mathcal{E}, \pi, f_{\text{TI}}, \gamma} \mathcal{W}_{f_{d_{\boldsymbol{x}}}}\left(\frac{1}{N} \sum_{i=1}^{N} \delta_{x_{i}}, \sum_{i_{1}, \dots, i_{Q}} \gamma\left(\left[\mathbf{e}_{i_{1}}, \dots, \mathbf{e}_{i_{Q}}\right]\right)\cdot \right. \\
\left. \delta_{f_{\text{TI}}\left(\left[\mathbf{e}_{i_{1}}, \dots, \mathbf{e}_{i_{Q}}\right]\right)}\right)
\end{aligned}
\end{equation}

By utilizing the Monge definition, we have
\begin{equation}\label{eq46}
\begin{aligned}
\mathcal{W}_{f_{d_{\boldsymbol{x}}}}\left(\frac{1}{N} \sum_{i=1}^{N} \delta_{x_{i}}, \sum_{i_{1}, \ldots, i_{Q}} \gamma\left(\left[\mathbf{e}_{i_{1}}, \ldots, \mathbf{e}_{i_{Q}}\right]\right) \delta_{f_{\text{TI}}\left(\left[\mathbf{e}_{i_{1}}, \ldots, \mathbf{e}_{i_{Q}}\right]\right)}\right) \\
= \min _{\mathcal{G}: \mathcal{G} \# \mathbb{P}_{\boldsymbol{x}}=f_{\text{TI}} \# \gamma} \mathbb{E}_{\boldsymbol{x} \sim \mathbb{P}_{\boldsymbol{x}}}\left[f_{d_{\boldsymbol{x}}}(\boldsymbol{x}, \mathcal{G}(\boldsymbol{x}))\right] \\
= \frac{1}{N} \min _{\mathcal{G}: \mathcal{G} \# \mathbb{P}_{\boldsymbol{x}}=f_{\text{TI}} \# \gamma} \sum_{i=1}^{N} f_{d_{\boldsymbol{x}}}\left(x_{i}, \mathcal{G}\left(x_{i}\right)\right),
\end{aligned}
\end{equation}
where $\mathcal{G}$ is a mapping function that maps the data point from the true distribution $\mathbb{P}_{\boldsymbol{x}}$ to the generative distribution points in $f_{\text{TI}} \# \gamma$. Thus, $\mathcal{G}\left(x_{n}\right)=f_{\text{TI}}\left(\left[\mathbf{e}_{i_{1}}, \ldots, \mathbf{e}_{i_{Q}}\right]\right)$ for some $i_{1}, \ldots, i_{Q}$. In addition, $\left|T^{-1}\left(f_{d}\left(\left[\mathbf{e}_{i_{1}}, \ldots, \mathbf{e}_{i_{M}}\right]\right)\right)\right|$, $k= 1, \ldots, K$ are proportional to $\gamma\left(\left[\mathbf{e}_{i_{1}}, \ldots, \mathbf{e}_{i_{Q}}\right]\right)$. 

Let $\sigma_{1}, \ldots, \sigma_{Q}$ be functions mapping $\{1, \ldots, N\} \rightarrow\{1, \ldots, K\}$, such that for
$\forall i = 1, \ldots, N$, 
$\mathcal{G}\left(x_{n}\right)=f_{\text{TI}}\left(\left[\mathbf{e}_{\sigma_{1}(n)}, \ldots, \mathbf{e}_{\sigma_{Q}(n)}\right]\right)$. Thus, we have a fixed allocation space $\sigma_{1}, \ldots, \sigma_{M} \in \Sigma_{\pi}$. It follows as
\begin{equation}\label{eq47}
\begin{aligned}
\mathcal{W}_{f_{d_{\boldsymbol{x}}}}\left(\frac{1}{N} \sum_{n=1}^{N} \delta_{x_{n}}, \sum_{i_{1}, \ldots, i_{Q}} \gamma\left(\left[\mathbf{e}_{i_{1}}, \ldots, \mathbf{e}_{i_{Q}}\right]\right) \delta_{f_{\text{TI}}\left(\left[\mathbf{e}_{i_{1}}, \ldots, \mathbf{e}_{i_{Q}}\right]\right)}\right)\\ = \frac{1}{N} \min _{\sigma_{1: Q} \in \Sigma_{\pi}} \sum_{i=1}^{N} f_{d_{\boldsymbol{x}}}\left(x_{i} , f_{\text{TI}}\left(\left[\mathbf{e}_{i_{1}}, \ldots, \mathbf{e}_{i_{Q}}\right]\right)\right).
\end{aligned}
\end{equation}

Further, the optimal solution of Eq.(\ref{eq45}) can rewrote as 
\begin{equation}\label{eq48}
\min _{f_{\text{TI}}} \min _{\mathcal{E}, \pi} \min _{\sigma_{1: Q} \in \Sigma_{\pi}} \sum_{i=1}^{N} f_{d_{\boldsymbol{x}}}\left(x_{i} , f_{\text{TI}}\left(\left[\mathbf{e}_{i_{1}}, \ldots, \mathbf{e}_{i_{Q}}\right]\right)\right),
\end{equation}
which directly implies Lemma \ref{Lemma1}, because the relationship between the $\left|\sigma_{q}^{-1}(k)\right|$ and the probability distribution $\gamma$ is
\begin{equation}\label{eq49}
\begin{aligned}
\left|\sigma_{q}^{-1}(k)\right| \propto \sum_{i_{1}, \ldots, i_{q}=k, \ldots, i_{Q}} \gamma\left(\left[\mathbf{e}_{i_{1}}, \ldots, \mathbf{e}_{i_{Q}}\right]\right)=\pi_{k}^{q}.
\end{aligned}
\end{equation}

Eventually, we complete the proof about Lemma \ref{Lemma1}.

\section{Proof of Theorem \ref{Theorem1}}

Given Lemma \ref{Lemma1}, we prove that Eq.(\ref{eq30}) is equivalent to
\begin{equation}\label{eq50}
\begin{aligned}
\min _{\mathcal{E}, \pi, f_{\text{TI}}, \gamma \in \Gamma} \min _{\bar{T}: \bar{T} \# \mathbb{P}_{\boldsymbol{x}} = \gamma} 
\mathbb{E}_{\boldsymbol{x} \sim \mathbb{P}_{\boldsymbol{x}},\left[\mathbf{e}_{i_{1}}, \ldots, \mathbf{e}_{i_{Q}}\right] \sim \bar{T}} \left[
f_{d_{\boldsymbol{x}}}\left(f_{\text{TI}}\left( \right. \right. \right. \\ \left. \left. \left.
\left[ \mathbf{e}_{i_{1}}, \ldots, \mathbf{e}_{i_{Q}}\right] \right), \boldsymbol{x} \right)\right],
\end{aligned}
\end{equation}
where $\bar{T}$ is a stochastic discrete encoder mapping data example $\boldsymbol{x}$ directly to the codebooks. To this end, we need to prove that
\begin{equation}\label{eq51}
\begin{aligned}
\mathcal{W}_{f_{d_{\boldsymbol{x}}}} & \left(f_{\text{TI}} \#\gamma,\mathbb{P}_{\boldsymbol{x}}\right)\\ &= \min_{\bar{T}:\bar{T}\#\mathbb{P}_{\boldsymbol{x}}=\gamma}\mathbb{E}_{\boldsymbol{x}\sim\mathbb{P}_{\boldsymbol{x}},\left[\mathbf{e}_{i_{1}}, \ldots, \mathbf{e}_{i_{Q}}\right] \sim \bar{T}} \left[f_{d_{\boldsymbol{x}}} \left(f_{\text{TI}}\left(\right. \right. \right. \\ &\left. \left. \left. \quad\quad\quad\quad\quad\quad\quad\quad\quad\quad
\left[\mathbf{e}_{i_{1}}, \ldots, \mathbf{e}_{i_{Q}}\right]\right), \boldsymbol{x} \right) \right].
\end{aligned}
\end{equation}

Let $\bar{T}$ be a stochastic discrete encoder such that $\bar{T} \# \mathbb{P}_{\boldsymbol{x}} = \gamma$ (i.e., $\boldsymbol{x} \sim \mathbb{P}_{\boldsymbol{x}}$ and $\left[\mathbf{e}_{i_{1}}, \ldots, \mathbf{e}_{i_{Q}}\right] \sim \bar{T}$ implies $\left[\mathbf{e}_{i_{1}}, \ldots, \mathbf{e}_{i_{Q}}\right] \sim \gamma$).
We consider $\mu_{d,\mathbf{e}}$ as the joint distribution of ($\boldsymbol{x}, \left[\mathbf{e}_{i_{1}}, \ldots, \mathbf{e}_{i_{Q}}\right]$), and also consider $\mu_{f_{\mathbf{e}}, d}$ as the joint distribution including $\left(\boldsymbol{x}, \boldsymbol{x}^{\prime}\right) \sim \mu_{f_{\mathbf{e}}, d}$ where $\boldsymbol{x}^{\prime}=f_{\text{TI}}\left(\left[ \mathbf{e}_{i_{1}}, \ldots, \mathbf{e}_{i_{Q}}\right] \right)$. This follows that $\mu_{f_{\mathbf{e}}, d} \in \Gamma\left(f_{\text{TI}} \# \gamma, \mathbb{P}_{\boldsymbol{x}}\right)$ which admits both as its marginal distribution have:
\begin{equation}\label{eq52}
\begin{aligned}
& \mathbb{E}_{\boldsymbol{x} \sim \mathbb{P}_{\boldsymbol{x}},\left[\mathbf{e}_{i_{1}}, \ldots, \mathbf{e}_{i_{Q}}\right] \sim \bar{T}(\boldsymbol{x})} \left[ f_{d_{\boldsymbol{x}}} \left(f_{\text{TI}}\left(\left[\mathbf{e}_{i_{1}}, \ldots, \mathbf{e}_{i_{Q}}\right]\right), \boldsymbol{x} \right) \right] 
\\ & = 
\mathbb{E}_{\left(\boldsymbol{x}, \left[\mathbf{e}_{i_{1}}, \ldots, \mathbf{e}_{i_{Q}}\right]\right) \sim \mu_{d,\mathbf{e}}} \left[ f_{d_{\boldsymbol{x}}} \left(f_{\text{TI}}\left(\left[\mathbf{e}_{i_{1}}, \ldots, \mathbf{e}_{i_{Q}}\right]\right), \boldsymbol{x} \right) \right]  \\
&\stackrel{(1)}{=}\mathbb{E}_{\left(\boldsymbol{x}, \boldsymbol{x}^{\prime}\right) \sim \mu_{f_{\mathbf{e}}, d}} \left[ f_{d_{\boldsymbol{x}}}\left(\boldsymbol{x}, \boldsymbol{x}^{\prime}\right) \right] \\
&\geq \min_{\mu_{f_{\mathbf{e}}, d} \in \Gamma\left(f_{\text{TI}} \# \gamma, \mathbb{P}_{\boldsymbol{x}}\right)} \mathbb{E}_{\left(\boldsymbol{x}, \boldsymbol{x}^{\prime}\right) \sim \mu_{f_{\mathbf{e}}, d}} \left[ f_{d_{\boldsymbol{x}}}\left(\boldsymbol{x}, \boldsymbol{x}^{\prime}\right) \right] \\
&= \mathcal{W}_{f_{d_{\boldsymbol{x}}}}(f_{\text{TI}} \# \mu, \mathbb{P}_{\boldsymbol{x}}).
\end{aligned}
\end{equation}

Note that we have the Eq.(\ref{eq30}) due to $\left(i d, f_{\text{TI}}\right) \# \mu_{d, \mathbf{e}}=\mu_{f_{\mathbf{e}}, d}$, which ensures its validity. Thus, we can obtain 
\begin{equation}\label{eq53}
\begin{aligned}
\min _{\bar{T}: \bar{T} \# \mathbb{P}_{\boldsymbol{x}} = \gamma} 
\mathbb{E}_{\boldsymbol{x} \sim \mathbb{P}_{\boldsymbol{x}},\left[\mathbf{e}_{i_{1}}, \ldots, \mathbf{e}_{i_{Q}}\right] \sim \bar{T}(\boldsymbol{x})} \left[
f_{d_{\boldsymbol{x}}}\left(f_{\text{TI}}\left( \right. \right. \right. \\ \left. \left. \left.
\left[ \mathbf{e}_{i_{1}}, \ldots, \mathbf{e}_{i_{Q}}\right] \right), \boldsymbol{x} \right)\right] \geq \mathcal{W}_{f_{d_{\boldsymbol{x}}}}\left(f_{\text{TI}} \# \gamma, \mathbb{P}_{x}\right).
\end{aligned}
\end{equation}

Also, take $\mu_{f_{\mathbf{e}}, \mathbf{e}} \in \Gamma\left(f_{\text{TI}} \# \gamma, \mathbb{P}_{\boldsymbol{x}}\right)$, which forms a deterministic coupling between the codewords $\left[\mathbf{e}_{i_{1}}, \ldots, \mathbf{e}_{i_{Q}}\right] \sim \gamma$ and the generated data $\boldsymbol{x}=f_{\text{TI}}\left(\left[ \mathbf{e}_{i_{1}}, \ldots, \mathbf{e}_{i_{Q}}\right] \right)$, imply $\left(\left[\mathbf{e}_{i_{1}} \ldots, \mathbf{e}_{i_{Q}}\right], \boldsymbol{x}\right) \sim \mu_{\mathbf{e}, f \mathbf{e}}$. Applying the gluing lemma (refer to Lemma 5.5 in \cite{Santambrogio2015OptimalTF}), there exists a joint distribution $\mu \in \Gamma\left(\gamma, f_{\text{TI}} \# \gamma, \mathbb{P}_{\boldsymbol{x}}\right)$ which admits $\mu_{f_{\mathbf{e}}, d}$ and $\mu_{f_{\mathbf{e}}, \mathbf{e}}$ as the corresponding joint
distributions. Denoting $\mu_{d,\mathbf{e}}\in\Gamma\left(\mathbb{P}_{\boldsymbol{x}},\gamma\right)$ as the marginal distribution of $\mu$ over $\mathbb{P}_{\boldsymbol{x}}$, $\gamma$, we then reach
\begin{equation}\label{eq54}
\begin{aligned}
&\mathbb{E}_{\left(\boldsymbol{x}, \boldsymbol{x}^{\prime}\right)\sim \mu_{f_{\mathbf{e}}, d}}\left[f_{d_{\boldsymbol{x}}}\left(\boldsymbol{x}, \boldsymbol{x}^{\prime}\right)\right]\\ & =\mathbb{E}_{\left(\left[\mathbf{e}_{i_{1}} \ldots, \mathbf{e}_{i_{Q}}\right], \boldsymbol{x}^{\prime}, \boldsymbol{x}\right)\sim \mu}\left[f_{d_{\boldsymbol{x}}}\left(\boldsymbol{x}, \boldsymbol{x}^{\prime}\right)\right]\\ & =\mathbb{E}_{\left(\left[\mathbf{e}_{i_{1}} \ldots, \mathbf{e}_{i_{Q}}\right], \boldsymbol{x}\right) \sim \mu_{d,\mathbf{e}}, \boldsymbol{x}^{\prime}=f_{\text{TI}}\left(\left[\mathbf{e}_{i_{1}} \ldots, \mathbf{e}_{i_{Q}}\right]\right)}\left[f_{d_{\boldsymbol{x}}}\left(\boldsymbol{x}, \boldsymbol{x}^{\prime}\right)\right] \\
&=\mathbb{E}_{\left(\left[\mathbf{e}_{i_{1}} \ldots, \mathbf{e}_{i_{Q}}\right], \boldsymbol{x}\right) \sim \mu_{d,\mathbf{e}}} \left[f_{d_{\boldsymbol{x}}} \left(f_{\text{TI}}\left(\left[\mathbf{e}_{i_{1}}, \ldots, \mathbf{e}_{i_{Q}}\right]\right), \boldsymbol{x} \right) \right] \\
&=\mathbb{E}_{\boldsymbol{x}\sim\mathbb{P}_{\boldsymbol{x}}, \left[\mathbf{e}_{i_{1}} \ldots, \mathbf{e}_{i_{Q}}\right] \sim \bar{T}} \left[f_{d_{\boldsymbol{x}}} \left(f_{\text{TI}}\left(\left[\mathbf{e}_{i_{1}}, \ldots, \mathbf{e}_{i_{Q}}\right]\right), \boldsymbol{x} \right) \right] \\
&\geq \min_{\bar{T}:\bar{T}\#\mathbb{P}_{\boldsymbol{x}}=\gamma}\mathbb{E}_{\boldsymbol{x}\sim\mathbb{P}_{\boldsymbol{x}},\left[\mathbf{e}_{i_{1}}, \ldots, \mathbf{e}_{i_{Q}}\right] \sim \bar{T}} \left[f_{d_{\boldsymbol{x}}} \left(f_{\text{TI}}\left(\right. \right. \right. \\ &\left. \left. \left. \quad\quad\quad\quad\quad\quad\quad\quad\quad\quad
\left[\mathbf{e}_{i_{1}}, \ldots, \mathbf{e}_{i_{Q}}\right]\right), \boldsymbol{x} \right) \right],
\end{aligned}
\end{equation}
where $\bar{T}=\mu_{d,\mathbf{e}}(\cdot\mid \boldsymbol{x})$. Thus, the OP in Eq.(\ref{eq30}) follows that
\begin{equation}\label{eq55}
\begin{aligned}
\mathcal{W}_{f_{d_{\boldsymbol{x}}}} & \left(f_{\text{TI}}\#\gamma,\mathbb{P}_{\boldsymbol{x}}\right) \\ & = \min_{\mu_{f_{\mathbf{e}}, d
}\in \Gamma \left(f_{\text{TI}}\#\gamma,\mathbb{P}_{\boldsymbol{x}}\right)} \mathbb{E}_{\left(\boldsymbol{x}, \boldsymbol{x}^{\prime}\right)\sim \mu_{f_{\mathbf{e}}, d}}\left[f_{d_{\boldsymbol{x}}}\left(\boldsymbol{x}, \boldsymbol{x}^{\prime}\right)\right] \\& \geq \min_{\bar{T}:\bar{T}\#\mathbb{P}_{\boldsymbol{x}}=\gamma}\mathbb{E}_{\boldsymbol{x}\sim\mathbb{P}_{\boldsymbol{x}},\left[\mathbf{e}_{i_{1}}, \ldots, \mathbf{e}_{i_{Q}}\right] \sim \bar{T}} \left[f_{d_{\boldsymbol{x}}} \left(f_{\text{TI}}\left(\right. \right. \right. \\ &\left. \left. \left. \quad\quad\quad\quad\quad\quad\quad\quad\quad\quad
\left[\mathbf{e}_{i_{1}}, \ldots, \mathbf{e}_{i_{Q}}\right]\right), \boldsymbol{x} \right) \right].
\end{aligned}
\end{equation}

This completes the proof for the equality in Eq.(\ref{eq51}), which means that Eq.(\ref{eq50}) satisfy the need condition. Thus, we further need to prove Eq.(\ref{eq50}) is equivalent to the limited Theorem \ref{Theorem1},
\begin{equation}\label{eq56}
\min_{\mathcal{E},\pi,\bar{f}_{\mathrm{TI}}}\min_{\gamma\in\Gamma}\min_{\bar{T}:\bar{T}\#\mathbb{P}_{\mathrm{z}}=\gamma} \mathbb{E}_{\boldsymbol{x} \sim \mathbb{P}_{\boldsymbol{x}}} \left[f_{d_{\boldsymbol{x}}}\left(f_{\text{TI}}\left(\bar{T}(\boldsymbol{x})\right),\boldsymbol{x}\right)\right],
\end{equation}
where $\bar{T}$ is a deterministic discrete encoder mapping data example directly to the codebooks. Note that the OP in Eq.(\ref{eq56}) is a special case of the OP in Eq.(\ref{eq50}) when we restrict to searching for deterministic discrete encoders.

Given the optimal solution $\mathcal{E}^{*1}$, $\pi^{*1}$, $\gamma^{*1}$, $f_{\text{TI}}^{*1}$, and $\bar{T}^{*1}$ of the OP in Eq.(\ref{eq50}), we indicate how to construct the optimal solution for the OP in Eq.(\ref{eq56}). Firstly, we construct $\mathcal{E}^{*2}=\mathcal{E}^{*1}$, $f_{\text{TI}}^{*2}=f_{\text{TI}}^{*1}$, and then denote $\bar{T}^{*2}\left(\boldsymbol{x}\right) = \mathrm{argmin}_{[\mathbf{e}_{i_{1}}\dots,\mathbf{e}_{i_{Q}}]}f_{d_{\boldsymbol{x}}}\left(f_{\text{TI}}^{*2}\left([\mathbf{e}_{i_{1}}\dots,\mathbf{e}_{i_{Q}}]\right),\boldsymbol{x}\right)$.

Thus, $\bar{T}^{*2}$ is a deterministic discrete encoder mapping data example directly to a sequence of codewords. Let us define $\pi_{k}^{*q2} = Pr\left(\bar{T}_{q}^{*2} (\boldsymbol{x})=\mathbf{e}_{k}:\boldsymbol{x}\sim\mathbb{P}_{\boldsymbol{x}}\right)$, where $\bar{T}^{*2} (\boldsymbol{x})=[\bar{T}_{q}^{*2}(\boldsymbol{x})]_{q=1}^{Q}$, meaning that $\bar{T}^{*2}\#\mathbb{P}_{\boldsymbol{x}}=\gamma^{*2}$, admitting $\mathbb{P}_{\mathbf{e}^{*2},\pi^{*q2}}$ as its marginal distributions. Thus, we reach 
\begin{equation}\label{eq57}
\begin{aligned}
&\mathbb{E}_{\boldsymbol{x} \sim \mathbb{P}_{\boldsymbol{x}}} \left[ f_{d_{\boldsymbol{x}}} \left( f_{\text{TI}}^{*2} \left( \bar{T}^{*2} \left( \boldsymbol{x} \right) \right), \boldsymbol{x} \right) \right] 
\\ & \leq 
\mathbb{E}_{\boldsymbol{x} \sim \mathbb{P}_{\boldsymbol{x}}, \left[ \mathbf{e}_{i_1}, \ldots, \mathbf{e}_{i_M} \right] \sim \bar{T}^{*1} (\boldsymbol{x})} \left[ f_{d_{\boldsymbol{x}}} \left( f_{\text{TI}}^{*1} \left( \left[ \mathbf{e}_{i_1}, \ldots, \mathbf{e}_{i_M} \right] \right), \boldsymbol{x} \right) \right].
\end{aligned}
\end{equation}

Moreover, because $\mathcal{E}^{*2}$, $\pi^{*2}$, $f_{\text{TI}}^{*2}$, and $\bar{T}^{*2}$ are also a feasible solution of Eq.(\ref{eq56}), we have 
\begin{equation}\label{eq58}
\begin{aligned}
&\mathbb{E}_{\boldsymbol{x} \sim \mathbb{P}_{\boldsymbol{x}}} \left[ f_{d_{\boldsymbol{x}}} \left( f_{\text{TI}}^{*2} \left( \bar{T}^{*2} \left( \boldsymbol{x} \right) \right), \boldsymbol{x} \right) \right] 
\\ & \geq 
\mathbb{E}_{\boldsymbol{x} \sim \mathbb{P}_{\boldsymbol{x}}, \left[ \mathbf{e}_{i_1}, \ldots, \mathbf{e}_{i_M} \right] \sim \bar{T}^{*1} (\boldsymbol{x})} \left[ f_{d_{\boldsymbol{x}}} \left( f_{\text{TI}}^{*1} \left( \left[ \mathbf{e}_{i_1}, \ldots, \mathbf{e}_{i_M} \right] \right), \boldsymbol{x} \right) \right].
\end{aligned}
\end{equation}

This means that
\begin{equation}\label{eq59}
\begin{aligned}
\mathbb{E}_{\boldsymbol{x} \sim \mathbb{P}_{\boldsymbol{x}}} \left[ f_{d_{\boldsymbol{x}}} \left( f_{\text{TI}}^{*2} \left( \bar{T}^{*2} \left( \boldsymbol{x} \right) \right), \boldsymbol{x} \right) \right] 
& = \\
\mathbb{E}_{\boldsymbol{x} \sim \mathbb{P}_{\boldsymbol{x}}, \left[ \mathbf{e}_{i_1}, \ldots, \mathbf{e}_{i_M} \right] \sim \bar{T}^{*1} (\boldsymbol{x})} 
& \left[ f_{d_{\boldsymbol{x}}} \left( f_{\text{TI}}^{*1} \left( \left[ \mathbf{e}_{i_1}, \ldots, \mathbf{e}_{i_M} \right] \right), \boldsymbol{x} \right) \right],
\end{aligned}
\end{equation}
and $\mathcal{E}^{*2}$, $\pi^{*2}$, $\gamma^{*2}$, $f_{\text{TI}}^{*2}$, and $\bar{T}^{*2}$ are also the optimal solution of Eq.(\ref{eq56}). Until this step, we complete this proof about Eq.(\ref{eq56}). At this time, we can extend to a more general form by directly utilizing the generalized task loss $L_{\text{task}}$ and the unified task inference module $\bar{f}_{\text{TI}}$, not just having decoders.

\section{Proof of Theorem \ref{Theorem2}}
To begin with, we need to prove the following lemma that is necessary for the proof of Theorem \ref{Theorem2}. 
\begin{lemma}\label{Lemma3}
Consider $\mathcal{E}$, $\pi$, $f_{\text{TI}}$, and $T$ as a feasible solution of the Eq.(\ref{eq33}). Denoting $\bar{T}^q(\boldsymbol{x}) = \mathrm{argmin}_{\mathbf{e}}\rho_z(T^q(\boldsymbol{x})),\mathbf{e})={\text{VQ}}_{\mathcal{E}}(\boldsymbol{x})$, then $\bar{T}^q(\boldsymbol{x})$ is a Borel measurable function and hence also $\bar{T}(\boldsymbol{x})=[\bar{T}^{q}(\boldsymbol{x})]_{q=1}^{Q}$.
\end{lemma}


The proof with respect to Lemma \ref{Lemma3} is shown below.

We denote the set $A_{k}$ on the latent space as
\begin{equation}\label{eq60}
\begin{aligned}
A_{k}=\{z:\rho_{z}(z,\mathbf{e}_{k})<\rho_{z}(z,\mathbf{e}_{j}),\forall j\neq k\}\\=\{z:{\text{VQ}}_{\mathcal{E}}(z)=\mathbf{e}_{k}\}.
\end{aligned}
\end{equation}

$A_{k}$ is known as a Voronoi cell w.r.t. the metric $\rho_z$. If we consider a continuous metric $\rho_z$, $A_{k}$ is a measurable set. Assuming a
Borel measurable function $B$, we need to prove that $\left(\bar{T}^{q}\right)^{-1}(B)$ is a Borel measurable set on the data space. Then, let $B\cap\{\mathbf{e}_1,..,\mathbf{e}_K\}=\{\mathbf{e}_{i_1},...,\mathbf{e}_{i_t}\}$, we prove that $\left(\bar{T}^{q}\right)^{-1}(B)=\cup_{j=1}^{t}\left(\bar{T}^{q}\right)^{-1}\left(A_{i_{j}}\right)$. 

Indeed, take
$\boldsymbol{x}\in(\bar{T}^{q})^{-1} (B)$, then $(\bar{T}^{q})^{-1}(\boldsymbol{x})\in B$, implying that $(\bar{T}^{q})^{-1}(\boldsymbol{x})={\text{VQ}}_{\mathcal{E}}(\boldsymbol{x})=\mathbf{e}_{i_j}$ for some $j=1,...,t$. This means that $T^q(\boldsymbol{x})\in A_{i_j}$ for some $j=1,...,t$. Therefore, we have $(\bar{T}^{q})^{-1} (B)\subset\cup_{j=1}^{t}(T^{q})^{-1}\left(A_{i_{j}}\right)$.

Given $\boldsymbol{x} \in \cup_{j=1}^{t}(T^{q})^{-1}\left(A_{i_{j}}\right)$, then we have $T^q(\boldsymbol{x})\in A_{i_j}$, hence $\bar{T}^q(\boldsymbol{x}) = {\text{VQ}}_{\mathcal{E}}(\boldsymbol{x}) = \mathbf{e}_{i_j}$ for some $j=1,...,t$. Thus, $\bar{T}^q(\boldsymbol{x})\subset B$ or equivalently $\boldsymbol{x}\in(\bar{T}^q(\boldsymbol{x}))^{-1} (B)$, implying $(\bar{T}^q(\boldsymbol{x}))^{-1} (B)\supset\cup_{j=1}^{t}(T^q(\boldsymbol{x}))^{-1}\left(A_{i_{j}}\right)$. 

Finally, we obtain $(\bar{T}^{q})^{-1}\left(B\right)=\cup_{j=1}^{t}(T^{q})^{-1}\left(A_{i_{j}}\right)$, which concludes our proof because $T^{q}$ is a measurable function and $A_{k}$ are measurable sets.

To proceed with, we prove the Theorem \ref{Theorem2} by the detailed process. Given the optimal solution $\mathcal{E}^{*1}$, $\pi^{*1}$, $\gamma^{*1}$, $f_{\text{TI}}^{*1}$, and $T^{*1}$ of Eq.(\ref{eq33}), we conduct the optimal solution for Eq.(\ref{eq32}). Let us conduct $\mathcal{E}^{*2}=\mathcal{E}^{*1}$, $f_{\text{TI}}^{*2}=f_{\text{TI}}^{*1}$, and define $\bar{T}^{*2}\left(\boldsymbol{x}\right)={\text{VQ}}_{{\mathcal{E}}^{*1}}\left(T^{*1}\left(\boldsymbol{x}\right)\right)={\text{VQ}}_{{\mathcal{E}}^{*2}}\left(T^{*1}\left(\boldsymbol{x}\right)\right)$. 
Thus, we prove that $\mathcal{E}^{*2}$, $\pi^{*2}$, $f_{\text{TI}}^{*2}$, and $\bar{T}^{*2}$ are optimal solution of the Eq.(\ref{eq32}). Define $\gamma^{*2}={\text{VQ}}_{{\mathcal{E}}^{*2}}\#(T^{*1}\#\mathbb{P}_{\boldsymbol{x}})$. By this definition, we yield $\bar{T}^{*2}\#\mathbb{P}_{\boldsymbol{x}}=\gamma^{*2}$ and hence
$\mathcal{W}_{f_{d_z}}\left(\bar{T}^{*2}\#\mathbb{P}_{\boldsymbol{x}},\gamma^{*2}\right)=0$. Then we need to verify two aspects for this proof process: 

$\text{(i)}$ $\bar{T}^{*2}$ is a Borel-measurable function.

$\text{(ii)}$ Given a feasible solution $\mathcal{E}$, $\pi$, $f_{\text{TI}}$, $\gamma$, $\bar{T}$ of Eq.(\ref{eq32}), we reach 
\begin{equation}\label{eq61}
\begin{aligned}
\mathbb{E}_{\boldsymbol{x} \sim \mathbb{P}_{\boldsymbol{x}}} \left[ f_{d_{\boldsymbol{x}}} \left( f_{\text{TI}}^{*2} \left( \bar{T}^{*2} \left( \boldsymbol{x} \right) \right), \boldsymbol{x} \right) \right] 
& \\\leq 
\mathbb{E}_{\boldsymbol{x} \sim \mathbb{P}_{\boldsymbol{x}}} 
& \left[f_{d_{\boldsymbol{x}}} \left( f_{\text{TI}} \left(\bar{T}(\boldsymbol{x})\right), \boldsymbol{x} \right) \right].
\end{aligned}
\end{equation}

Regarding (i), it is a direct conclusion because the application of Lemma \ref{Lemma3} to $\mathcal{E}^{*1}$, $\pi^{*1}$, $f_{\text{TI}}^{*1}$, and $T^{*1}$.

Then, for (ii), we can further derive as 
\begin{equation}\label{eq62}
\begin{aligned}
&\mathbb{E}_{\boldsymbol{x}\sim\mathbb{P}_{\boldsymbol{x}}}\left[f_{d_{\boldsymbol{x}}}\left(f_{\text{TI}}^{* 2}\left(\bar{T}^{* 2}\left(\boldsymbol{x}\right)\right),\boldsymbol{x}\right)\right]+\lambda\mathcal{W}_{f_{d_z}}\left(\bar{T}^{*2}\#\mathbb{P}_{\boldsymbol{x}},\gamma^{*2}\right) \\
&=\mathbb{E}_{\boldsymbol{x}\sim\mathbb{P}_{\boldsymbol{x}}}\left[f_{d_{\boldsymbol{x}}}\left(f_{\text{TI}}^{*2}\left(\bar{T}^{*2}\left(\boldsymbol{x}\right)\right),\boldsymbol{x}\right)\right] \\
&=\mathbb{E}_{\boldsymbol{x}\sim\mathbb{P}_{\boldsymbol{x}}}\left[f_{d_{\boldsymbol{x}}}\left(f_{\text{TI}}^{*1}\left({\text{VQ}}_{{\mathcal{E}}^{*2}}\left(T^{*1}\left(\boldsymbol{x}\right)\right)\right),\boldsymbol{x}\right)\right] \\
&=\mathbb{E}_{\boldsymbol{x}\sim\mathbb{P}_{\boldsymbol{x}}}\left[f_{d_{\boldsymbol{x}}}\left(f_{\text{TI}}^{*1}\left({\text{VQ}}_{{\mathcal{E}}^{*1}}\left(T^{*1}\left(\boldsymbol{x}\right)\right)\right),\boldsymbol{x}\right)\right] \\
&\leq\mathbb{E}_{\boldsymbol{x}\sim\mathbb{P}_{\boldsymbol{x}}}\left[f_{d_{\boldsymbol{x}}}\left(f_{\text{TI}}^{*1}\left({\text{VQ}}_{{\mathcal{E}}^{*1}}\left(T^{*1}\left(\boldsymbol{x}\right)\right)\right),\boldsymbol{x}\right)\right]\\
& \quad \quad \quad \quad \quad \quad \quad \quad +\lambda\mathcal{W}_{f_{d_z}}\left(T^{*1}\#\mathbb{P}_{\boldsymbol{x}},\gamma^{*1}\right).
\end{aligned}
\end{equation}

Due to $\bar{T}\#\mathbb{P}_{\boldsymbol{x}}=\gamma$ which is a discrete distribution over $\mathcal{E}^{Q}$, we obtain ${\text{VQ}}_{\mathcal{E}}(\bar{T}(\boldsymbol{x}))=\bar{T}(\boldsymbol{x})$. Note that $\mathcal{E}$, $\pi$, $f_{\text{TI}}$, and $\bar{T}$ is also a feasible solution of Eq.(\ref{eq33}) because $\bar{T}$ is also a specific encoder mapping from data space to the latent space,
we have
\begin{equation}\label{eq63}
\begin{aligned}
&\mathbb{E}_{\boldsymbol{x}\sim\mathbb{P}_{\boldsymbol{x}}}\left[f_{d_{\boldsymbol{x}}}\left(f_{\text{TI}}\left({\text{VQ}}_{\mathcal{E}}\left(\bar{T}\left(\boldsymbol{x}\right)\right)\right),\boldsymbol{x}\right)\right]+\lambda\mathcal{W}_{f_{d_z}}\left(\bar{T}\#\mathbb{P}_{\boldsymbol{x}},\gamma\right)\\& \quad\quad\quad\quad
\geq\mathbb{E}_{\boldsymbol{x}\sim\mathbb{P}_{\boldsymbol{x}}}\left[f_{d_{\boldsymbol{x}}}\left(f_{\text{TI}}^{*1}\left({\text{VQ}}_{{\mathcal{E}}^{*1}}\left(\bar{T}^{*1}\left(\boldsymbol{x}\right)\right),\boldsymbol{x}\right)\right)\right] \\ &\quad\quad\quad\quad\quad\quad\quad\quad\quad\quad\quad\quad +  \lambda\mathcal{W}_{f_{d_z}}\left(\bar{T}^{*1}\#\mathbb{P}_{\boldsymbol{x}},\gamma^{*1}\right).
\end{aligned}
\end{equation}

Based on above conditions, we arrive at 
\begin{equation}\label{eq64}
\begin{aligned}
&\mathbb{E}_{\boldsymbol{x}\sim\mathbb{P}_{\boldsymbol{x}}}\left[f_{d_{\boldsymbol{x}}}\left(f_{\text{TI}}\left(\bar{T}\left(\boldsymbol{x}\right)\right),\boldsymbol{x}\right)\right]\\&\quad\quad\quad\quad \geq\mathbb{E}_{\boldsymbol{x}\sim\mathbb{P}_{\boldsymbol{x}}}\left[f_{d_{\boldsymbol{x}}}\left(f_{\text{TI}}^{*1}\left({\text{VQ}}_{{\mathcal{E}}^{*1}}\left(T^{*1}\left(\boldsymbol{x}\right)\right)\right),\boldsymbol{x}\right)\right]\\ &\quad\quad\quad\quad\quad\quad\quad\quad\quad\quad\quad\quad +\lambda\mathcal{W}_{f_{d_z}}\left(\bar{T}^{*1}\#\mathbb{P}_{\boldsymbol{x}},\gamma^{*1}\right).
\end{aligned}
\end{equation}

Combining Eq.(\ref{eq62}) with Eq.(\ref{eq64}), we obtain inequality (ii). This concludes our proof. As described in Appendix B, we also convert Theory \ref{Theorem2} into a generalized objective framework.

\section{Proof of Lemma \ref{Lemma2}}
Considering the conditions proved in the  above appendix, we denote $\mu^{*q}\in\Gamma\left(T^{q}\#\mathbb{P}_{\boldsymbol{x}},\mathbb{P}_{\mathbf{e},\pi^{q}}\right)$ the optimal coupling for the WS distance $\mathcal{W}_{\rho_{z}} (T^{q}\#\mathbb{P}_{\boldsymbol{x}},\mathbb{P}_{\mathbf{e},\pi^{q}})$. Then, we construct a coupling $\mu\in\Gamma(T\#\mathbb{P}_{\boldsymbol{x}},\gamma)$.
Firstly, we sample $X\sim\mathbb{P}_{\boldsymbol{x}}$ and sample $\mathcal{E}_{q} \sim \mu^{*q}(\cdot | T^{q}(X)), q=1, \ldots, Q$. Let $\gamma^{*}$ be the law of $\mathcal{E}^{1}, \ldots, \mathcal{E}^{Q}$ and $\mu_{*}$ be the law of $(T(X),\left[\mathcal{E}^{1}, \ldots, \mathcal{E}^{Q}\right])$. Next, we define $\pi^{*q}$ such that $\mathbb{P}_{\mathbf{e},\pi^{*q}}$ is the marginal distribution of $\gamma^{*}$ over $\mathcal{E}_{q}$. And we have $\gamma^*\in\Gamma(\mathbb{P}_{\mathbf{e},\pi^1},\ldots,\mathbb{P}_{\mathbf{e},\pi^Q})$ and
$\mu^{*}\in\Gamma(T\#\mathbb{P}_{\boldsymbol{x}},\gamma^{*})$. It follow that
\begin{equation}\label{eq65}
\begin{aligned}
&\mathcal{W}_{f_{d_z}} (T\#\mathbb{P}_{\boldsymbol{x}},\gamma^{*}) 
\\&=\mathbb{E}_{(T(X),\left[\mathcal{E}^{1}, \ldots, \mathcal{E}^{Q}\right])\sim\mu^{*}}\left[f_{d_z}\left(\left[T^{1}\left(X\right),\ldots,T^{Q}\left(X\right)\right], \left[ \right. \right. \right. \\ &\left. \left. \left. \quad\quad\quad\quad\quad\quad\quad\quad\quad\quad\quad\quad\quad\quad\quad\quad\quad \mathcal{E}^{1}, \ldots, \mathcal{E}^{Q}\right]\right)\right] \\
&=\frac{1}{Q}\sum_{q=1}^{Q}\mathbb{E}_{\left(T^{q}(X),\mathcal{E}_{q}\right)\sim\mu^{*q}}\left[\rho_{z}\left(T^{q}\left(X\right),\mathcal{E}_{q}\right)\right] \\
&=\frac{1}{M}\sum_{q=1}^{Q}\mathcal{W}_{\rho_{z}}\left(T^{q}\#\mathbb{P}_{\boldsymbol{x}},\mathbb{P}_{\mathbf{e},\pi^{q}}\right).
\end{aligned}
\end{equation}

Thus, we further obtain
\begin{equation}\label{eq66}
\begin{aligned}
&\min_{\pi}\min_{\gamma\in\Gamma}\mathcal{W}_{f_{d_z}}\left(T\#\mathbb{P}_{\boldsymbol{x}},\gamma\right)\leq\mathcal{W}_{f_{d_z}}\left(T\#\mathbb{P}_{\boldsymbol{x}},\gamma^*\right)\\&=\frac{1}{Q}\sum_{q=1}^{Q}\mathcal{W}_{\rho_z}\left(T^q\#\mathbb{P}_{\boldsymbol{x}},\mathbb{P}_{\mathbf{e},\pi^q}\right).
\end{aligned}
\end{equation}

Then, we complete the proof about Lemma \ref{Lemma2}.

\section{Proof of Corollary \ref{corollary1}:}

By the Monge definition, we can reach
\begin{equation}\label{eq67}
\begin{aligned}
&\mathcal{W}_{\rho_{z}}\left(T^{q}\#\mathbb{P}_{\boldsymbol{x}},\mathbb{P}_{\mathbf{e},\pi^{q}}\right)=\mathcal{W}_{\rho_{z}}\left(\frac{1}{N}\sum_{n=1}^{N}\delta_{T^{q}\left(x_{n}\right)},\sum_{k=1}^{K}\pi_{k}^{q}\delta_{\mathbf{e}_{k}}\right)\\&=\min_{\mathcal{G}:\mathcal{G}\#\left(T^{q}\#\mathbb{P}_{\mathbf{e}}\right)=\mathbb{P}_{\mathbf{e},\pi^{q}}}\mathbb{E}_{\boldsymbol{z}_{e}\thicksim T^{q}\#\mathbb{P}_{\boldsymbol{x}}}\left[\rho_{z}\left(\boldsymbol{z}_{e},\mathcal{G}\left(\boldsymbol{z}_{e}\right)\right)\right]\\&=\frac{1}{N}\min_{\mathcal{G}:\mathcal{G}\#\left(T^{q}\#\mathbb{P}_{\boldsymbol{x}}\right)=\mathbb{P}_{\mathbf{e},\pi}q}\sum_{n=1}^{N}\rho_{z}\left(T^{q}\left(x_{n}\right),\mathcal{G}\left(T^{q}\left(x_{n}\right)\right)\right).
\end{aligned}
\end{equation}

Since $\mathcal{G}\#\left(T^{q}\#\mathbb{P}_{\boldsymbol{x}}\right) = \mathbb{P}_{\mathbf{e},\pi^{q}}$, $\mathcal{G}\left(T\left(x_{n}\right)\right)=\mathbf{e}_{k}$ for some $k$. In addition, $\left|\mathcal{G}^{-1}\left(\mathbf{e}_{k}\right)\right|$ are proportional $\pi_{k}^q$. Then, we denote $\sigma:\{1,...,N\} \to \{1,\ldots,K\}$ such that $\mathcal{G}\left(T^{q}\left(x_{n}\right)\right)=\mathbf{e}_{\sigma(n)}$, we have $\sigma\in\Sigma_{\pi}$. It also follows that 
\begin{equation}\label{eq68}
\begin{aligned}
\mathcal{W}_{\rho_z} & \left(\frac{1}{N}\sum_{n=1}^{N}\delta_{T^q(x_n)},\sum_{k=1}^{K}\pi_k^q \delta_{\mathbf{e}_k}\right)\\& \quad\quad\quad\quad\quad=\frac{1}{N}\min_{\sigma\in\Sigma_\pi}\sum_{n=1}^{N}\rho_z\left(T^q\left(x_n\right),\mathbf{e}_{\sigma(n)}\right). 
\end{aligned}
\end{equation}

Finally, we complete the proof about Corollary \ref{corollary1} to provide the addition about this WS-Based codebook strategy.

\end{appendices}

\bibliographystyle{IEEEtran} 
\bibliography{bib}

\end{document}